\documentclass[11pt]{article}
 \pdfoutput=1
\usepackage{amsmath}
\usepackage{amssymb}
\usepackage{graphicx}
\usepackage{epsfig}
\usepackage{pdfpages}

\begin{document}

\newtheorem{Thm}{Theorem}[section]
\newtheorem{Cor}[Thm]{Corollary}
\newtheorem{Lem}[Thm]{Lemma}
\newtheorem{Prop}[Thm]{Proposition}
\newtheorem{Def}[Thm]{Definition}
\newtheorem{rem}[Thm]{Remark}
\newtheorem{Assump}[Thm]{Assumption}

 \title{\textbf{Optimal $k$-thresholding Algorithms for Sparse Optimization Problems}}
\author{YUN-BIN ZHAO\thanks{School of Mathematics, University of Birmingham,
Edgbaston, Birmingham B15 2TT,  United Kingdom ({\tt
y.zhao.2@bham.ac.uk}). }}

\date{(First version: 5 Oct 2018; Second version: 20 June 2019)}

\maketitle

\begin{abstract} The simulations indicate that the existing hard thresholding technique independent of the residual function may cause a  dramatic increase or numerical oscillation of the residual.  This inherit drawback of the hard thresholding renders the traditional thresholding algorithms  unstable and thus generally inefficient for solving practical sparse optimization problems.     How to overcome this weakness and develop a truly efficient thresholding method is a fundamental question in this field. The aim of this paper is to address this question by proposing a new thresholding technique based on the notion of optimal $k$-thresholding. The central idea for this new development is to connect the $k$-thresholding directly to the residual reduction during the course of algorithms.
    This leads to a natural design principle for the efficient thresholding methods. Under the restricted isometry property (RIP), we prove that the optimal thresholding based algorithms are globally convergent to the solution of sparse optimization problems.   The numerical experiments demonstrate  that when solving sparse optimization problems, the traditional hard thresholding methods have been significantly transcended by the proposed  algorithms which can even outperform the classic $\ell_1$-minimization method in many situations.\\
\end{abstract}



\section{Introduction}

  Let $A\in \mathbb{R}^{m\times n} ~ (m<n)$ be a given matrix, $ y\in \mathbb{R}^m $ be a given vector and $\varepsilon \geq 0$ be a given parameter.   Let  $\|x\|_0$ denote the  `$\ell_0$-norm' counting the number of nonzero entries of the vector $x \in \mathbb{R}^n. $
The  sparse optimization problem is to find a sparse (or the sparsest) vector, denoted by $x^*,$  such that $Ax^*$ can best fit the vector $y.$ This problem can be formulated as the minimization problem with a sparsity constraint
\ \begin{equation} \label{L0} \min_{x} \{\|Ax-y\|_2^2:  \|x\|_0 \leq k\},\end{equation}
where $k$ is a prescribed integer number, or formulated as the so-called  $\ell_0$-minimization problem
\begin{equation}\label {M2} \min_{x} \{\|x\|_0:  \|Ax-y\|_2\leq \varepsilon \}.  \end{equation}
Both (\ref{L0}) and (\ref{M2}) are the central  models for sparse signal recovery and   sparse representation of data on their redundant bases. These models provide an essential basis for the development of the theory and algorithms for compressed sensing  (see, e.g., \cite {BDE09, E10, EK12, FR13, Z18}). The problem  (\ref{L0}) has also been widely used  in the fields of statistical regressions and wireless communications (see, e.g.,  \cite{M02, BKM16, LYSM17}).

The problems (\ref{L0}) and (\ref{M2}) are NP-hard  in general \cite{N95}. The plausible algorithms for such problems can be briefly categorized into the following classes:   (i) Convex optimization methods (e.g., $\ell_1$-minimization \cite{CDS98},  reweighed $\ell_1$-minimization \cite{CWB08, FL09, ZL12}, and  dual-density-based reweighted $\ell_1$-minimization \cite{Z18, ZK15, ZL17});  (ii) heuristic methods (such as matching pursuit  \cite{MZ93}, orthogonal matching pursuit  \cite{MDZ94,TG07}, compressive sampling matching pursuit  \cite{NT09}, and subspace pursuit  \cite{DM09});   (iii) thresholding methods (e.g., soft thresholding  \cite{DDM04, D95,  E06}, hard thresholding  \cite{B12, BD08, BD09, F11}, graded hard thresholding pursuits  \cite{B14,  BFH16}, and the `firm' thresholding \cite{VW13}); (iv) integer programming methods \cite{BKM16}.

The use of thresholding techniques for signal denoising problems can be dated back to the seminal paper by Donoho and Johnstone \cite{DJ94}. Since then, various thresholding algorithms
 were  proposed  for sparse recovery or sparse approximation (see, e.g., Reeves and Kingsbury \cite{RK02}, Kingsbury and Reeves \cite{KR03},  Figueiredo and  Nowak \cite{FN03},  Starck et al. \cite{SNM03},  Herrity et al.  \cite{HGT06},  Blumensath and Davies \cite{BD08, BD09, BD10}, and Beck and Teboulle \cite{BT09}).     The thresholding  algorithms can be derived from different perspectives such as minimizing certain surrogate functions related to the residual function $\|y-Ax\|_2^2$ (see, e.g., \cite{BD08, DDM04,  L16}) and the necessary optimality conditions for minimization with sparsity constraints \cite{BE13, BE13b}. The  algorithms can be classified as soft thresholdings or hard thresholdings according to the nature of  thresholding operators.  The  soft ones are closely related to  the optimality condition of certain convex optimization (e.g., \cite{E06, VW13}) and have been widely analyzed in the literature (e.g., \cite{DDM04, E06,   FR08, HGT06, VW13}).  The hard thresholding ones for compressed sensing were analyzed  by Blumensath and Davies \cite{BD08, BD09, BD10}, Foucart \cite{F12, F11}, and Foucart and Rauhut \cite{FR13}.

 For convenience of discussion, we focus on the problem (\ref{L0}) in this paper.  Given $z\in \mathbb{R}^n,$  let ${\cal H}_k(z) $ denote the vector obtained by retaining  the $k$ largest  magnitudes of  $z $ and zeroing out the remaining entries of $z.$  The operator $ {\cal H}_k(\cdot) $ is referred to as the \emph{hard thresholding operator}. Since the $k$ largest magnitudes of $z$ may not be unique (see Theorem \ref{Thm-0201} for details),  ${\cal H}_k(z)$ might contain more than one vectors in some situations.     The  iterative hard thresholding (IHT) algorithm takes the scheme
\begin{equation} \label{IHT} x^{p+1} \in {\cal H}_k \left(x^p+  \tau A^T(y-Ax^p) \right) \end{equation}
  to search the solution of  (\ref{L0}), where  $A^T$ is the transpose of $A$ and $\tau>0$ is a stepsize which can be iteratively updated or a fixed number (such as $\tau \equiv 1$).
The iterative scheme (\ref{IHT}) can be dated back to Landweber \cite{L51}. The Landweber iteration $z^{p+1} = z^p+  \tau A^T(y-Az^p) $ is  essentially the gradient method for minimizing the  function $  \|y-Ax\|_2^2. $ Thus an intuitive idea for possibly solving the  problem (\ref{L0}) is to perform the hard thresholding on the Landweber iteration, leading to the iterative scheme (\ref{IHT}).

 The analyses  in \cite{BD08, BD09,  F12, F11, M09} show that the convergence of the IHT algorithm can be guaranteed under the restricted isometry property (RIP) or a mutual coherence condition. The RIP was first introduced by Cand\`es and Tao \cite{CT05} (see also Cand\`es \cite{C08}) to study the signal recovery via the $\ell_1$-minimization method. However, the  empirical evidences indicate that the  efficiency of the IHT is actually low.
 For instance, taking
 $ A= \left[
        \begin{array}{cccc}
          1 & 2 & 3 & 4\\
          5 & 6 & 7 & 8 \\
        \end{array}
      \right] $ and $y=   \left[\begin{array}{c}
                                     1 \\
                                     5 \\
                                   \end{array}
                                 \right],
$  it is evident that $x^*= (1,0,0,0)^T $ is the solution to the problem (\ref{L0}).  However, the  IHT starting from  $x^0  =0 $ generates the following sequence:  $
~  x^1  =   {\cal H}_1(u^0) = (0, 0, 0, 44),
~ x^2  =    {\cal H}_1(u^1)= (0, 0, 0, -3432), ~ x^3  =    {\cal H}_1(u^2)= (0, 0, 0, -271170),  \dots  , $ where $u^p: = x^p+A^T (y-Ax^p).$   The sequence $\{x^p\}$ diverges, and the corresponding sequence of  $  r(x^p)=\|y-A x^p\|_2$ (i.e.,  $ r(x^0)=\sqrt{26},  ~  r(x^1)= 388.6309, ~  r(x^2) =3.0702e+04, ~ r(x^3)= 2.4254e+06, \dots  $)   also diverges so quickly.  Thus there is a  huge gap between the theoretical efficiency and practical performance of the IHT. This stimulates the study of various acceleration and stabilization techniques for
this sort of algorithms.

 The first idea for acceleration is using a stepsize as in (\ref{IHT}).
The algorithm with a fixed stepsize was called gradient descent with sparsification in \cite{GK09}. See also \cite{BE13,  B12, C11}. With iteratively updated stepsizes, the algorithm is called the normalized iterative hard thresholding in \cite{BD10}.
See also the so-called CGIHT algorithm in \cite{BTW15}.  Another idea is to minimize the residual over the support determined by the hard thresholding. With this idea, Foucart \cite{F11} proposed the following algorithm called hard thresholding pursuit (HTP):
\begin{equation} \label{(HTP1)}
   S^{p+1} = \textrm{supp} (\hat{z}),  ~ \hat{z} \in   {\cal H}_k( x^p+ A^T(y-Ax^p)),
\end{equation}
\begin{equation}  \label{(HTP2)}
x^{p+1}  \in    \textrm{arg}\min_{x}\left \{\|y-Ax\|_2^2:  ~  \textrm{supp} (x) \subseteq S^{p+1}\right\}.
\end{equation}
 The step (\ref{(HTP2)})  is used to chase a better vector than $\hat{z}$  that can best fit  the vector $y.$ This idea is also used  in  compressive sensing matching pursuit  proposed by Needell and Tropp  \cite{NT09} and  in subspace pursuit  proposed by Dai and Milenkovic \cite{DM09}. As a generalization of the HTP, the graded hard thresholding pursuit  \cite{B14,  BFH16}    combines the step (\ref{(HTP2)}) and orthogonal matching pursuit. Other acceleration versions of  the IHT based on Nestrov's techniques \cite{N13} can be found in \cite{C11, KK17, KC14}.

In many situations, however, directly using the  operator ${\cal H}_k(\cdot)$ is not attractive from the perspective of the residual   $\|y-Ax\|_2^2 .$    The thresholding step (\ref{(HTP1)})  is actually independent of the residual reduction (see Section 3 for details) in the sense that it does not include any mechanism to reduce  the residual  in the course of iterations. It actually causes the divergence of the IHT in numerous situations, or significantly slows down the convergence of the algorithm.  Even aided with (\ref{(HTP2)}), numerical experiments demonstrate that   the values of the residual at the iterates generated by the HTP may still oscillate dramatically, rendering the algorithm inefficient in many situations.  Such an oscillation phenomenon (see Fig. \ref{Fig-A} (a) in Section 5)  was caused by the hard thresholding operator  which often increases instead of decreasing the residual. To our knowledge, the  existing  ideas for acceleration do not serve the purpose of eliminating such an inherent drawback of  the operator ${\cal H}_k.$

In this paper, retaining $k$ entries of a vector and zeroing out its remaining entries is referred to as a $k$-thresholding of the vector.  Motivated by the above observation,  we explore the following idea in order to develop efficient thresholding methods:  \emph{The $k$-thresholding should be performed to serve for the purpose of  residual reduction.}   Linking the thresholding with  residual reduction enables us to introduce the notion of \emph{optimal $k$-thresholding}. More specifically, it enables us to select a set of $k$ entries of a vector that achieves the least residual  among all possible selections of $k$ entries. Clearly, the optimal $k$ entries is not necessarily the $ k$  largest  magnitudes of the vector.  Based on this notion, we propose the optimal $k$-thresholding (OT) algorithm and the optimal $k$-thresholding pursuit (OTP). Since the subproblems in OT and OTP are  binary quadratic minimization problems which are usually not convenient to solve directly, we propose the   relaxed optimal $k$-thresholding (ROT)   and the relaxed optimal $k$-thresholding pursuit  (ROTP)   which  naturally result from the tightest convex relaxation of the binary optimization problem in OT and OTP. The ROTP and its further enhanced versions (ROTP2 and ROTP3) turn out to be a    new and powerful  generation of  thresholding algorithms which significantly reverse the adversity of using the traditional hard thresholding.

The OT and OTP algorithms  are shown to have the guaranteed success for sparse signal recovery under the   RIP bound $\delta_{2k}< 0.5349 $ (see Theorem \ref{Thm-OIHT-01}  for details).  This bound is largely theoretical  by assuming that  the binary subproblems in OT or OTP can be successfully solved by certain methods.  The guaranteed success of  the  ROT and ROTP  is also proved in this paper under the RIP bound $\delta_{3k} <1/5.$  The empirical results collected from random examples of sparse optimization problems show that the ROTP and its enhanced versions remarkably outperform the IHT and HTP as anticipated, and the ROTP2 and ROTP3 are efficient enough to outperform the $\ell_1$-minimization in numerous situations (see Section 5 for details).  Simulations also demonstrate that the proposed algorithms are stable in the sense that the residual is steadily  reduced during the course of iterations.

The paper is organized as follows.  Section 2 provides some notations, definitions and properties of the hard thresholding operator.   The new thresholding methods  are described in Section 3.  The theoretical performance of several proposed algorithms are rigorously shown under the RIP  condition in Section 4. Numerical results for the ROTP and its enhanced versions are reported in Section 5.

\section{Preliminary}

\subsection{Notation} $\mathbb{R}^n$ denotes the $n$-dimensional Euclidean space, and $ \mathbb{R}^{m\times n}$ stands for the set of $m\times n $ metrics. The set of $n$-dimensional binary vectors is denoted by $\{0,1\}^n. $  All vectors are column vectors unless otherwise specified.  We use $\textbf{\textrm{e}} $ to denote the vector of ones and $I$ to denote the identity matrix. For a vector $x\in \mathbb{R}^n,$ $\|x\|_2,$ $\|x\|_1$ and $\|x\|_\infty$ denote the $\ell_2$-, $\ell_1$- and $\ell_\infty$-norms, respectively, and $|x|$ denotes the absolute vector of $x$, i.e., $|x|_i = |x_i|$ for $i=1, \cdots, n. $    The support of  $x$ is denoted by $\textrm{supp}(x) $ which is the index set $ \{i:x_ i\not =0\}. $ The nonnegative vector $x$ is written as $x\geq 0. $ For two vectors $x $ and $ y, $  the inequality  $x\geq y$  means $x-y$ is a nonnegative vector.    Given a set $S\subseteq \{1,2,\dots, n\},$
$|S|$ denotes the cardinality of $S$ and   $\overline{S}=  \{1, 2,\dots, n\} \backslash S $ denotes the complement set of $S. $   Given $x\in \mathbb{R}^n$ and   $S\subseteq \{1, \dots, n\},$   $x_S\in \mathbb{R}^n $   denotes the vector obtained by retaining the components of $x$ indexed by $S$ and zeroing out the remaining components of $x.$  That is, for every $ i=1, \dots, n,$  $(x_S)_i= x_i $ if $i\in S; $ otherwise, $(x_S)_i =0. $    For $x, y \in  \mathbb{R}^n$,  the vector $x \otimes y$ is the Hadamard product of $x$ and $y,$  i.e., $x\otimes y =(x_1y_1, \dots, x_n y_n)^T. $  A vector is said to be $k$-sparse if $\|x\|_0\leq k. $

\subsection{Characteristics of hard thresholding operator ${\cal H}_k (\cdot)$}
Given an integer number $k,$
   a vector  $  w\in \{0,1\}^n $  with exactly $k$ nonzero entries can be represented as $w\in \{0,1\}^n$ and $\textrm{\textbf{e}}^T w =k.$
Denote the set of such vectors by \begin{equation} \label{WWW-set}  {\cal W}^{(k)} = \left\{w:  ~ w\in \{0,1\}^n, ~ \textrm{\textbf{e}}^T w = k\right\}. \end{equation} Note that   the Hadamard product
 $ \widetilde{z} = z \otimes w ,$ where  $ z \in \mathbb{R}^n $ and $w \in {\cal W}^{(k)}, $ is the vector thresholded from $z$  by retaining $z_i$ corresponding to $ w_i=1$ and zeroing out the remaining ones. We introduce the following definition.

\begin{Def}  Given $z\in \mathbb{R}^n$ and $w \in {\cal W}^{(k)}, $ the vector $\widetilde{z} = z \otimes w $   is called a $k$-thresholding vector of $z,$ and the associated vector $w\in {\cal W}^{(k)} $ is called a $k$-thresholding indicator. If the $k$-thresholding retains the  $k$ largest magnitudes of  $z, $  it is referred to as the  hard $k$-thresholding of $z. $
\end{Def}

Clearly,  ${\cal H}_k(z)$ is the set of $z \otimes w^\#, $  where $ w^\# \in {\cal W}^{(k)} $ is an indicator for   the  $k$   largest magnitudes of  $ z. $ Denote by ${\cal W}^*(z) \subseteq {\cal W}^{(k)} $  be the set of indicators for the $k$ largest magnitudes of $z.$  Then ${\cal W}^*(z)=  \{ w^\#\in {\cal W}^{(k)}:  ~  z \otimes w^\# \in {\cal H}_k(z) \} . $
We also note that  $ {\cal H}_k (z) $ is the set of vectors which are the best $k$-term approximation of  $z, $   namely,
 $$ {\cal H}_k (z) = \textrm{arg} \min_{u}    \{ \|z-u\|_1:  ~  \|u\|_0 \leq k\}.$$ Denote by $\sigma_k(z)_1$ the error of  the best $k$-term approximation of $z, $ i.e., $$\sigma_k(z)_1  = \min_{u}   \{ \|z-u\|_1:  ~ \|u\|_0 \leq k \}. $$   Clearly, $\sigma_k(z)_1 =0$ if and only if $z$ is $k$-sparse. In this paper,  $z$ is said to be $k$-compressible if $\sigma_k(z)_1 $ is small.
Note that
 $ \sigma_k(z)_1= \|z-\hat{z}\|_1  ~ \textrm{ for any } \hat{z}  \in  {\cal H}_k(z) . $
By the definition of ${\cal W}^*(z)$ and ${\cal W}^{(k)}$, we see that   for any  $  w^\# \in {\cal W}^*(z)$
\begin{equation} \label{sigma-01} \sigma_k(z)_1 =   \|z- (z\otimes w^\# )\|_1 = \|z\otimes (\textrm{\textbf{e}}-w^\#) \|_1 = |z|^T (\textrm{\textbf{e}}-w^\#) .  \end{equation}
 Since $ \| z \otimes w \|_0 \leq k $   for any $w\in {\cal W}^{(k)}, $     by  the definition of  $ \sigma_k(z)_1$, we have
\begin{equation} \label{sigma-02}  \sigma_k(z)_1  \leq   \|z- (z\otimes w)  \|_1  = |z|^T (\textrm{\textbf{e}}-w) ~ \textrm{ for any }  w \in  {\cal W}^{(k)} . \end{equation}
It follows from (\ref{sigma-01}) and (\ref{sigma-02})  that every hard $k$-thresholding indicator $w^\# \in  {\cal W}^*(z)  $ is exactly  the solution to the following 0-1  integer programming problem:
\begin{equation} \label{0-1-IP}  \min_{w}  \left\{  |z|^T (\textrm{\textbf{e}}-w) :     ~ \textrm{\textbf{e}}^T w =k,  ~  w  \in   \{0,1\}^n \right \}.  \end{equation}
  This    problem is  very easy to solve via the linear programming (LP) relaxation
\begin{equation} \label{LP-relaxation}   \min_{w}  \left\{  |z|^T(\textrm{\textbf{e}}- w) :    ~ \textrm{\textbf{e}}^T w =k,  ~ 0\leq w  \leq  \textbf{\textrm{e}}  \right\},  \end{equation}
 as indicated by the following Lemma.

 \begin{Lem} \label{Lem-0201} Given   $z\in \mathbb{R}^n , $   let $\hat{\gamma} (z) $ be the optimal objective value of (\ref{LP-relaxation}) and let $\widehat{S} $ be the set of optimal solutions of (\ref{LP-relaxation}) that are extreme points of the feasible set. Then  $\hat{\gamma} (z)=\sigma_k(z)_1$  and $\widehat{S} =  {\cal W}^*(z). $   Thus  $w^\# \in \widehat{S} $
if and only if $   z \otimes  w^\# \in {\cal H}_k(z). $
\end{Lem}

\emph{Proof.}   Consider the feasible set of the problem (\ref{LP-relaxation})
\begin{equation} \label{LP-feasible-set}  \mathcal{P} : =   \{w \in \mathbb{R}^n:  ~ \textrm{\textbf{e}}^T w = k, ~  0\leq w \leq  \textbf{\textrm{e}} \}.   \end{equation}
Let $V$ denote the set of extreme points of this polyhedron. By introducing the nonnegative variable $u \in \mathbb{R}^n,$ the linear system in $ \mathcal{P} $ can be written as
$ \textrm{\textbf{e}}^T w = k,  ~ w+u =\textrm{\textbf{e}}, ~ w\geq 0 $ and $  u\geq 0,$ that is,
$   \left[
      \begin{array}{cc}
        \textrm{\textbf{e}}^T  & 0 \\
        I & I \\
      \end{array}
    \right] \left[
              \begin{array}{c}
                w \\
                u \\
              \end{array}
            \right] = \left[
                        \begin{array}{c}
                          k \\
                          \textrm{\textbf{e}} \\
                        \end{array}
                      \right]$ and $  \left[
              \begin{array}{c}
                w \\
                u \\
              \end{array}
            \right] \geq 0,  $
where $I$ is the $n\times n$ identity matrix. Note that the matrix  $\left[
      \begin{array}{cc}
        \textrm{\textbf{e}}^T  & 0 \\
        I & I \\
      \end{array}
    \right] $ is totally unimodular, and the right-hand-side vector $ \left[
                        \begin{array}{c}
                          k \\
                          \textrm{\textbf{e}} \\
                        \end{array}
                      \right]  $ of the above system is an integer vector. The total-unimodularity theory implies that  every extreme point of the polyhedron ${\cal P} $ is an integer vector. Therefore, by the structure of ${\cal P},$  every extreme point of ${\cal P}$ must be a binary vector with $k$ entries being ones. This means  $V\subseteq {\cal W}^{(k)}. $
 By the LP theory,  at least one of the extreme points of  ${\cal P} $ must be optimal.  Thus  $ \emptyset \not = \widehat{S} \subseteq V \subseteq {\cal W}^{(k)} . $    It  follows from (\ref{sigma-02}) that $\sigma_k(z)_1 \leq \hat{\gamma} (z)= |z|^T (\textrm{\textbf{e}} - \hat{w}) $ for $\hat{w} \in \widehat{S} \subseteq {\cal W}^{(k)}. $  Let $\tilde{w} \in {\cal W}^*(z) $  which is contained in the feasible set of (\ref{LP-relaxation}).
By optimality  and  (\ref{sigma-01}), we have $ \hat{\gamma}(z) \leq |z|^T(\textrm{\textbf{e}}- \tilde{w}) =  \sigma_k(z)_1. $ Therefore $\sigma_k(z)_1= \hat{\gamma}(z),$ from which it is not difficult to see that ${\cal W}^*(z)$ is exactly the set of optimal  solutions of  (\ref{LP-relaxation}) that are extreme points of the feasible set, and hence $\widehat{S} = {\cal W}^*(z). $ Therefore  $w^\# \in \widehat{S} $
if and only if $ z \otimes  w^\# \in {\cal H}_k(z). $

                     \vskip 0.07in

It is well known that finding the hard $k$-thresholding of a vector  is very easy and can be done in several ways. The equivalence of (\ref{0-1-IP}) and  (\ref{LP-relaxation}) implies that solving the LP problem (\ref{LP-relaxation}) is an alternative way.
We now point out that the condition for ${\cal H}_k(z)$ being a singleton can be completely characterized.  Denote by $z^*$ the  nonincreasing rearrangement of $|z|$, i.e., $z_1^*\geq z_2^* \geq  \dots \geq z_n^* \geq 0,$ and $\pi$ is a permutation of $\{1, \dots, n\} $ such that $z^*_j= |z_{\pi(j)}| $ for $j=1, \dots, n. $ The following theorem claims that
 $ {\cal H}_k (z)$ is  a singleton if and only if the   $k$th largest absolute entry of $z$ is strictly larger than the  $(k+1)$th largest absolute entry.

\begin{Thm}\label{Thm-0201} Let $z\in \mathbb{R}^n $ be a given vector and $z^*$ be the nonincreasing arrangement of $|z|. $  The following three statements are equivalent:
    \emph{(a)}   $ {\cal H}_k(z) $  is a singleton;
   \emph{(b)} The solution of the LP problem (\ref{LP-relaxation}) is unique;
 \emph{ (c)}  $ z^*_k >    z^*_{k+1}.  $
\end{Thm}

\emph{Proof.}  The equivalence of (a) and (b)  follows from Lemma \ref{Lem-0201} straightaway.  It is sufficient to show  the equivalence of (c) and (a).  First we note that when $ z^*_k= z^*_{k+1}$,  there are at least two distinct sets of the $k$ largest magnitudes of $z$, so ${\cal H}_k(z) $ is not unique.  Thus (a) implies (c).   We now show that  (c) also implies (a).  Assume that $ z^*_k > z^*_{k+1}$ and denote by the set
$L_k(z) = \{ \pi(i):  ~  |z_{\pi(i)}|= z^*_i,  ~ i=1, \dots, k\} , $ which  is the  set of indices for the $k$ largest magnitudes of $z.$
   Let  $w^* $ be an arbitrary optimal   solution of (\ref{LP-relaxation}) which is an extreme point of its feasible set.
Note that $$
\sum_{i\notin L_k(z) }  |z_i|   w^*_i \leq \left[\max_{i\notin L_k(z)} |z_i|\right]
 \sum_{i\notin L_k(z)}   w^*_i  = z^*_{k+1}  \left[k-\sum_{i\in L_k(z)}
 w^*_i\right] =  z^*_{k+1} \sum_{i\in L_k(z)}  (1- w^*_i). $$
From Lemma \ref{Lem-0201}, we have $    \sigma_k(z)_1=|z|^T (\textrm{\textbf{e}}-w^*) . $ This together with the  inequality above implies
\begin{eqnarray*}  \sigma_k(z)_1  &    = &  \sum_{i\in L_k(z)}  |z_i| (1-w^*_i)   +  \sum_{i\notin L_k(z)} |z_i| (1-w^*_i) \\
 & = &  \sum_{i\notin L_k(z)} |z_i| + \sum_{i\in L_k(z)}  |z_i| (1-w^*_i) - \sum_{i\notin L_k(z)} |z_i| w^*_i\\
 & \geq &  \sigma_k(z)_1 + \sum_{i\in L_k(z)}  |z_i| (1-w^*_i)   - z^*_{k+1} \sum_{i\in L_k(z)}   (1-w^*_i)\\
 & = & \sigma_k(z)_1 + \sum_{i\in L_k(z)}  (|z_i|- z^*_{k+1} )    (1- w^*_i) .
 \end{eqnarray*} With the fact $0\leq w^* \leq \textbf{\textrm{e}}$ and   $ |z_i| > z^*_{k+1} $ for every  $ i \in L_k(z), $ the inequality above implies that   $w^*_i =1 $ for all $ i\in L_k(z).  $ By the constraints of $(\ref{LP-relaxation}),$ the remaining $n-k$ components of $ w^*$ are equal to $ 0.   $  So $w^*$   is uniquely determined. This  means  the set of the optimal   solutions of (\ref{LP-relaxation}) which are extreme points of its feasible set  contains only a single vector, and thus $ {\cal H}_k (z)$ is unique (by Lemma \ref{Lem-0201}).

\vskip 0.07in

 The link between ${\cal H}_k, $ ${\cal P} $   and $ \sigma_k(\cdot)_1$ indicates that  performing  ${\cal H}_k (\cdot)$ on a vector is nothing but minimizing the error of the $k$-term approximation of the vector, which  is independent of the residual function $\|y-Az\|_2^2.$ This motivates us to consider a new thresholding strategy in the next section.

\section{Optimal $k$-thresholding algorithms and their relaxations}

   The classic steepest descent method for minimizing the residual $\|y-Ax\|_2^2 $ is deeply rooted in the following theoretical basis: When the current iterate  is not a minimizer of the  function, moving from the iterate in the direction of negative gradient  of the function (with a certain stepsize  if necessary)  leads to the decrease in  the value of this function.   This theoretical basis, however,  is generally lost when the  operator ${\cal H}_k(\cdot)$ is applied to the vector $u^p : = x^p + A^T(y-Ax^p). $ As we have pointed out in  Section 2, the selection of the $k$ largest  magnitudes of this vector    is independent of the residual $\|y-Ax\|_2^2.$  Thus the hard $k$-thresholding may cause the increase of the residual at  $\hat{u} \in {\cal H}_k (u^p),$  i.e., $\|y -A \hat{u}\|_2 >  \|y-A x^p\|_2. $ This is the main reason for the iterative scheme  $x^{p+1}\in  {\cal H}_k(u^p)$  being unstable and inefficient for solving sparse optimization problems,
 unless $u^p$ is  $k$-compressible (in which case $ {\cal H}_k\left(u^p\right) \approx u^p$) so that  the  scheme $  x^{p+1} \in  {\cal H}_k\left(u^p\right) $  is close to the   steepest descent method.

To  overcome the drawback of the  hard thresholding, we may link the $k$-thresholding with a residual function, and perform thresholding and residual reduction   simultaneously.
      This stimulates the following thresholding of a  vector $z \in \mathbb{R}^n :$
\begin{equation}
\label{Z-PR-01}  \alpha^*(u):= \min_{w}  \{\| y- A (u  \otimes w) \|^2_2:   ~  \textrm{\textbf{e}}^T w = k, ~  w \in \{0,1\}^n\} .
\end{equation}    In this model, performing a $k$-thresholding  of $u$ is  directly related to the residual function.  The  $k$-thresholding of $u$ resulting from (\ref{Z-PR-01}) admits the least residual, and thus it is better than  other $k$-thresholdings of $u,$ including ${\cal H}_k(u). $
 We use $w^*(u)$ to denote the optimal solution of  (\ref{Z-PR-01}).    Clearly,  the solution $w^*(u) $   relies on the choice of the objective function, which
 may take  other forms different from the one in (\ref{Z-PR-01}).  For instance, we may  minimize the $\ell_1 $-norm of the gradient of  $\| y- A x  \|_2^2, $   leading to the following  model:
 $$   \min _{w}   \left\{  \| A^T (y- A (u \otimes w))\|_1 :
  ~ \textrm{\textbf{e}}^T w=  k ,
  ~ w \in \{0,1\}^n  \right\}.
$$
  For simplicity, however,  we only focus on the model (\ref{Z-PR-01})  and its convex relaxations in this paper.   We  introduce the following definition.

\begin{Def} \label{Def0201} Given  $u \in \mathbb{R}^n, $  the solution of  (\ref{Z-PR-01}), denoted by $w^*(u), $   is called the optimal $k$-thresholding indicator, and the vector $  u  \otimes  w^* (u)  $ is called the optimal $k$-thresholding of $u .$  The operator $$Z^{\#}_k (u): = \{ u  \otimes  w^* (u) : ~ w^* (u) \textrm{ is an optimal solution of   (\ref{Z-PR-01})} \}  $$    is called the optimal $k$-thresholding operator.
\end{Def}

The solution of  (\ref{Z-PR-01}) may not be unique, and thus $Z^{\#}_k(u)$ might contain more than one vector. Since $ \alpha^*(u)=\|y-A v\|_2^2   $ for any $ v \in Z^{\#}_k (u),$ we may simply write this as  $\alpha^*(u)=\|y-A Z^{\#}_k (u)\|_2^2   $ no matter $ Z^{\#}_k (u)$ is a singleton or not.  By optimality,   we have
 \begin{equation} \label{OPT-01} \left \| y- A  Z^{\#}_k(u) \right \|_2  \leq  \| y- A  (u\otimes w) \|_2  ~ \textrm{ for any } w \in {\cal W}^{(k)},       \end{equation}
where ${\cal W}^{(k)} $ is given in (\ref{WWW-set}). This implies that   \begin{equation}  \label{OPT-02}  \left\| y- A Z^{\#}_k(u)  \right \|_2  \leq  \min_{\hat{u}\in {\cal H}_k(u) } \| y- A\hat{u} \|_2 . \end{equation}  Thus the optimal $k$-thresholding is never worse than the hard $k$-thresholding from the perspective of residual reduction.
In terms of optimal $k$-thresholding, we obtain the following iterative scheme:
\begin{equation} \label{IOHT}  x^{p+1}\in   Z^\#_k\left( x^p+ A^T (y-Ax^p) \right). \end{equation}
This method is referred to as the optimal $k$-thresholding (OT) algorithm, which by  the definition of $Z^{\#}_k(\cdot)  $  is described explicitly as follows.

\vskip 0.08in

\textbf{OT Algorithm:}  Input $(A, y,k).$ Give an initial point $x^0\in \mathbb{R}^n$ and  repeat the following steps  until a stoping criterion is satisfied:
\begin{itemize}
 \item[S1.] At  $x^p$, set $u^p  = x^p + A^T (y-A x^p) $ and solve the problem
  \begin{equation}   \label {HT-QP}   \min_{w}   \left\{ \|y- A (u^p  \otimes   w) \|^2_2 : ~  \textrm{\textbf{e}}^T w= k , ~   w \in  \{0,1\}^n \right\}.
  \end{equation}
 \item[S2.] Let $w^*(u^p)$ be the  solution to the problem (\ref{HT-QP}), and set
 $$ x^{p+1}: =  u^p  \otimes w^*(u^p) . $$

  \end{itemize}

In general,  the input vector $u^p$ in S1 is not  $k$-sparse, but the output $u^p  \otimes w^*(u^p)$ of S1 is a compressed (in fact, $k$-sparse) vector.   So the step S1 above can be called a ``compressing step". The binary optimization problem (\ref {HT-QP}) is known to be NP-hard \cite{CAP08} (see also \cite{BT18}).     This problem is similar to the best subset selection model in statistics \cite{M02}, and Bertsimas et al. \cite{BKM16} developed a mixed-integer optimization formulation to deal with similar binary   optimization problems. Their study indicates that in many cases the problem like (\ref{HT-QP}) can be directly solved by exploiting the integer programming structure, and thus it might not be always necessary to consider a convex relaxation of the problem (see the numerical results in \cite{BKM16} for more details).

  In this paper, however, we focus on the convex relaxation of  the binary problem  (\ref{HT-QP}). The convex relaxation turns out be a very efficient technique for the development of practical thresholding algorithms based on the above OT framework. To relax the problem (\ref{HT-QP}), an immediate idea  is to replace the binary constraint $w\in \{0,1\}^n$  with  the simple restriction $w\in [0,1]^n.$  In other words, we replace the feasible set $  {\cal W}^{(k)} $ of (\ref{HT-QP}) with the polytope $ {\cal P}$  defined in (\ref {LP-feasible-set}).  From the  proof of Lemma \ref{Lem-0201},  we see that  ${\cal P} $ is the convex hull, i.e.,  the tightest convex relaxation of ${\cal W}^{(k)}. $
 This leads to the following convex relaxation counterpart of (\ref{HT-QP}):
\begin{equation}   \label {HT-QP-Relax}  \gamma^*(u^p): = \min_{w}   \left\{  \|y- A (u^p  \otimes   w) \|^2_2 : ~  \textrm{\textbf{e}}^T w= k , ~  0\leq w\leq \textrm{\textbf{e}} \right\},
  \end{equation}
which can be solved efficiently  by interior-point methods or other optimization methods. Let $w^p$ be the solution of (\ref{HT-QP-Relax}). Since $w^p$  may not be exactly $k$-sparse, we  apply ${\cal H}_k$ to the vector $u^p\otimes w^p$ to produce the next $k$-sparse iterate. This leads to following relaxed optimal $k$-thresholding method termed the `ROT' algorithm.

 \vskip 0.08in

 \textbf{ROT Algorithm:} Input $(A, y, k). $ Give an initial point $x^0$ and repeat the following steps until a stoping criterion is satisfied:

 \begin{itemize}
 \item[S1.] At  $x^p$, set  $u^p  = x^p + A^T (y-A x^p) $  and solve the convex optimization problem  (\ref{HT-QP-Relax}) to obtain $w^p. $
\item [S2.]  Set
 $$ x^{p+1}\in  {\cal H}_k (u^p  \otimes  w^p).  $$
  \end{itemize}

 The first step above  can  still be called a ``compressing step" since the output $ u^p\otimes w^p$ is more compressible than $u^p$ in the sense that $\sigma_k( u^p  \otimes  w^p)_1 \leq \sigma_k (u^p)_1 $ which follows from the fact $0 \leq w^p \leq \textbf{\textrm{e}}. $ In fact, for a given vector $z\in \mathbb{R}^n,$   $\sigma_k (z)_1  $ is the sum of the $n-k$ smallest components of $|z|.$ Let $\Lambda $ denote the index set of the $n-k$ smallest components of $|u^p|. $ Then $\sigma_k (u^p)_1 =  \|(u^p)_\Lambda\|_1.  $ Therefore,
  $$\sigma_k( u^p  \otimes  w^p)_1 \leq \|(u^p  \otimes  w^p)_\Lambda\|_1   \leq \|(w^p)_\Lambda\|_\infty  \|(u^p)_\Lambda\|_1  \leq \sigma_k (u^p)_1,$$  where the last inequality follows from the fact  $ \|(w^p)_\Lambda\|_\infty  \leq 1 . $
  When  $k\ll n$ (which is typical in compressed sensing scenarios),  most components of   $w^p$    are very small and thus $\sigma_k( u^p  \otimes  w^p)_1$ might be much smaller than $\sigma_k(u^p)_1. $ In particular, $\sigma_k( u^p  \otimes  w^p)_1 =0  $ when $w^p \in    {\cal W}^{(k)}.  $
  The difference between the ROT and traditional hard thresholding methods is obvious.  Traditional ones directly apply the hard $k$-thresholding to $u^p$ without making any effort  to reduce the residual. When $u^p $ is not compressible, the hard thresholding $ {\cal H}_k (u^p) $  might dramatically raise the value of the residual function, causing  divergence or very slow convergence of the  iterates.  By contrast,   the ROT improves the efficiency of thresholdings by simultaneously compressing the vector $u^p$ and decreasing the residual. The ROT  integrates these two efforts  to overcome the drawback of performing ${\cal H}_k (\cdot)$ directly onto   non-compressible vectors.
We now point out an advantage of applying $ {\cal H}_k (\cdot)$ to a  compressible vector.

\begin{Lem}\label{Lem-03-01} Let  $u$ be an arbitrary vector in $\mathbb{R}^n.$  Then for any $\hat{u} \in {\cal H}_k(u), $   \begin{equation}\label{COM-ERROR}
\left| \|y-A \hat{u}\|_2^2 - \|y-A u \|_2^2 \right| \leq 2\|A^T(y-A u)\|_\infty   \sigma_k (u)_1 + \lambda_{\max} (A^TA)  (\sigma_k (u)_1)^2. \end{equation}
\end{Lem}

\emph{Proof.}   Let $\hat{u} \in {\cal H}_k(u).$ Note that
$$ \|y-A \hat{u}\|_2^2  =  \|y-A u \|_2^2 + 2[A^T(y-Au)]^T (\hat{u}-u) +  (\hat{u}-u)^T A^TA (\hat{u}-u).  $$
Thus,  $$ \left| \|y-A\hat{u}\|_2^2 - \|y-A u \|_2^2 \right| \leq 2\|A^T(y-Au) \|_\infty \|\hat{u}-u\|_1 +  \lambda_{\max} (A^TA) \|\hat{u}-u\|_2^2, $$
which together with
$ \|\hat{u}-u\|_2 \leq \|\hat{u}-u\|_1 = \sigma_k (u)_1   $ implies the  inequality (\ref{COM-ERROR}).

\vskip 0.08in

This lemma shows that if $\sigma_k (u)_1   $ is    small (i.e., $u$ is   $k$-compressible),  then $\|y-A \hat{u}\|_2^2  \approx \|y-A u \|_2^2 $ for any $\hat{u} \in {\cal H}_k(u). $  Thus performing a hard $k$-thresholding on  a compressible vector  will not dramatically raise the value of  the residual.
Since the output, $u^p\otimes w^p, $ of the first step  of ROT  is more compressible than the input vector $u^p,$   the way for generating $x^{p+1}$ in ROT is believed to be  more sensible than the way in IHT and HTP.
The next result interprets further why a hard $k$-thresholding should  apply to compressible vectors instead of non-compressible ones.

\begin{Thm} \label{Prop-03-01}   Let  $x^p \in \mathbb{R}^n $ be given and $u^p= x^p+ A^T (y-Ax^p). $ Let $\gamma^* (u^p)$  and $w^p$  be the optimal value and the  optimal solution of (\ref{HT-QP-Relax}), respectively, and let  $x^{p+1}\in {\cal H}_k(u^p\otimes w^p) . $   Denote by  $\alpha^* (u^p)  $    the optimal value of (\ref{HT-QP}).   Then the following two statements hold: \emph{(i)}   $\gamma^* (u^p) \leq \alpha^* (u^p)  \leq  \min_{\hat{u} \in {\cal H}_k(u^p)} \|y-A \hat{u} \|_2^2 ; $
\emph{(ii) } $  \| y-A x^{p+1} \|_2^2  \leq  \alpha^* (u^p)    $ provided that
$$ \sigma_k( u^p \otimes w^p )_1 \leq  \frac{  \sqrt{  \varphi(u^p,w^p) ^2 + 4 (\alpha^*(u^p) -\gamma^*(u^p)) \lambda_{\max}(A^TA) }- \varphi(u^p,w^p)    }
 { 2\lambda_{\max}(A^TA) }, $$  where
$ \varphi(u^p,w^p) = 2 \|A^T(y-A (u^p \otimes w^p ))\|_\infty.$
\end{Thm}

\emph{Proof.}   The statement (i) is obvious, following  directly from (\ref{OPT-02})  and the optimality of $w^p.$  Let  $\varphi(u^p,w^p) $ be defined as above.
 Consider the following quadratic function (in variable $t   $):
$ \vartheta (t) = \gamma^*(u^p)  + \varphi(u^p,w^p)   t +  \lambda_{\max}(A^TA) t^2 .  $ By  Lemma \ref{Lem-03-01},
\begin{eqnarray} \label{bound01}     \| y-A x^{p+1} \|^2_2
&    \leq  & \| y-A  (u^p  \otimes w^p )   \|_2^2  + 2 \|A^T(y-A (u^p  \otimes w^p )) \|_\infty  \sigma_k( u^p \otimes w^p )_1 \nonumber \\
 & &  ~~~ +  \lambda_{\max}(A^TA) (\sigma_k \left( u^p \otimes w^p )_1 \right)^2  \nonumber \\
&  = & \gamma^* (u^p)+ \varphi (u^p,w^p)  \sigma_k( u^p \otimes w^p )_1 +   \lambda_{\max}(A^TA) (\sigma_k( u^p \otimes w^p )_1 )^2 \nonumber \\
 &    = &  \vartheta ( \sigma_k( u^p \otimes w^p )_1).
\end{eqnarray}
It is easy to verify that
$ \vartheta (t) \leq \alpha^* (u^p)  $  provided that $ t$ is smaller than or equal to the following root of the quadratic equation $  \vartheta (t) =\alpha^* (u^p) :$
$$  \Omega(u^p, w^p) := \frac{ - \varphi(u^p,w^p) + \sqrt{  \varphi(u^p,w^p)^2 + 4 (\alpha^*(u^p)-\gamma^*(u^p)) \lambda_{\max}(A^TA) } } {2 \lambda_{\max}(A^TA) }. $$
Thus when
$\sigma_k( u^p \otimes w^p )_1 \leq \Omega(u^p, w^p),  $  we must have   $ \vartheta (\sigma_k( u^p \otimes w^p )_1) \leq \alpha^*(u^p).$ This, combined with  (\ref{bound01}),  implies that
$ \| y-A x^{p+1} \|^2_2  \leq  \alpha^*(u^p)  . $

\vskip 0.07in

This result shows  that if  $ \sigma_k( u^p \otimes w^p )_1 $ is small enough, then   $$ \max_{\bar{u} \in {\cal H}_k( u^p \otimes w^p)}  \| y-A \bar{u} \|_2 \leq   \|y-AZ^{\#}_k(u^p)\|_2 \leq  \min_{\hat{u} \in {\cal H}_k(u^p)}\|y-A \hat{u}\|_2,  $$ which means the iterates generated by the ROT  will never worse than the traditional hard thresholding algorithms from the perspective of  residual reductions.
The OT and ROT algorithms can be further enhanced by using the pursuit step (\ref{(HTP2)}).   The OT combined with (\ref{(HTP2)}) is referred to as the optimal $k$-thresholding pursuit (OTP), and the ROT algorithm combined with (\ref{(HTP2)}) is called the relaxed optimal $k$-thresholding pursuit (ROTP), which are described respectively as follows.

  \vskip 0.07in

  \textbf{OTP Algorithm:}  Input $(A, y,k).$ Given an initial point $x^0\in \mathbb{R}^n,$  repeat the following steps  until a stoping criterion is satisfied:
\begin{itemize}
 \item[S1.]  At  $x^p$, set $u^p  = x^p + A^T (y-A x^p) $ and solve the binary optimization problem (\ref{HT-QP}),
 let  $w^*(u^p)$ be a solution of  this problem.

\item[S2.] Set $ S^{p+1}: = \textrm{supp} (u^p  \otimes w^*(u^p) )  $ and
let $x^{p+1}$ be a solution to the problem
 $$ \min_{x} \{ \|y-A x\|_2^2:  ~ \textrm{supp} (x)\subseteq S^{p+1} \}. $$
  \end{itemize}

   \textbf{ROTP Algorithm:} Input $(A, y, k).$ Given an initial point $x^0\in \mathbb{R}^n, $  repeat the following steps until a  stoping criterion is satisfied:
 \begin{itemize}
 \item[S1. ] At  $x^p$, set $u^p = x^p + A^T (y-A x^p) $ and solve the convex optimization  problem
  $$  \min_{w}  \{  \| y- A  (u^p  \otimes w) \|_2^2:   ~~ \textrm{\textbf{e}}^T w= k, ~
   0\leq w \leq \textrm{\textbf{e} } \}
$$
 to generate a solution $w^p $ of this problem.

 \item[S2. ] Let  $ v \in   {\cal H}_k (u^p  \otimes  w^p) , ~~  S^{p+1} = \textrm{supp} (v), $
 and let  $x^{p+1}$ be a solution to the problem
 $$ \min_{x} \{ \|y-A x\|_2^2:  ~ \textrm{supp} (x)\subseteq S^{p+1} \}. $$
  \end{itemize}

 The vector $w^p$ generated by the  ``compressing step" of the ROTP might not be sparse enough.    This motivates  the following enhanced versions of the ROTP called ROTP2 and  ROTP3 which perform twice and three times of compressions of the data $u^p,$ respectively, before the operator ${\cal H}_k$ is  applied to the resulting compressible vector.    As shown by numerical experiments (see Section 5 for details), the aforementioned drawback  of the hard thresholding  will be remarkably overcame through compressing $u^p$ more than once.

\vskip 0.07in

  \textbf{ROTP2 Algorithm:} Input $(A, y, k).$ Given an initial point $x^0\in R^n, $  repeat the following steps until a stoping criterion is satisfied:
 \begin{itemize}
 \item[S1. ] At  $x^p$, set $u^p  = x^p + A^T (y-A x^p) $ and solve the   problem
  $$  \min_{w}  \{  \| y- A  (u^p  \otimes w) \|_2^2:   ~~ \textrm{\textbf{e}}^T w= k, ~
   0\leq w \leq \textrm{\textbf{e} } \}
$$
 to get  a solution $w^{(1)}$ to this problem. Then solve  the problem
$$  \min_{w}  \{  \| y- A  (u^p  \otimes w^{(1)} \otimes w) \|_2^2:   ~~ \textrm{\textbf{e}}^T w= k, ~
   0\leq w \leq \textrm{\textbf{e} } \}
$$ to get a solution $ w^{(2)}$ to this problem.

 \item[S2. ] Let $ v \in  {\cal H}_k (u^p  \otimes  w^{(1)} \otimes w^{(2)}) $ and $ S^{p+1}  = \textrm{supp}(v).  $
Let  $x^{p+1}$ be a solution to the problem
 $$ \min_{x} \{ \|y-A x\|_2^2:  ~ \textrm{supp} (x)\subseteq S^{p+1} \}. $$
  \end{itemize}

  \textbf{ROTP3 Algorithm:} Input $(A, y, k).$ Given an initial point $x^0\in R^n, $  repeat the following steps until a  stoping criterion is satisfied:
 \begin{itemize}
 \item[S1. ] At  $x^p$, set $u^p = x^p + A^T (y-A x^p) $ and solve the   problem
  $$  \min_{w}  \{  \| y- A  (u^p  \otimes w) \|_2^2:   ~~ \textrm{\textbf{e}}^T w= k, ~
   0\leq w \leq \textrm{\textbf{e} } \}
$$
 to get a solution  $w^{(1)}.$ Then solve
$$  \min_{w}  \{  \| y- A  (u^p \otimes w^{(1)} \otimes w) \|_2^2:   ~~ \textrm{\textbf{e}}^T w= k, ~
   0\leq w \leq \textrm{\textbf{e} } \}
$$
to obtain a solution $w^{(2)}$, and then solve
$$  \min_{w}  \{  \| y- A  (u^p \otimes w^{(1)}\otimes w^{(2)} \otimes w) \|_2^2:   ~~ \textrm{\textbf{e}}^T w= k, ~
   0\leq w \leq \textrm{\textbf{e} } \}
$$  to obtain a solution $w^{(3)}.$

 \item[S2. ] Let $ v \in {\cal H}_k (u^p  \otimes  w^{(1)} \otimes w^{(2)} \otimes w^{(3)})   $  and $ S^{p+1} = \textrm{supp}  (v) .$
 Let $x^{p+1}$ be the solution to the problem
 $$ \min_{x} \{ \|y-A x\|_2^2:  ~ \textrm{supp} (x)\subseteq S^{p+1} \}. $$
  \end{itemize}

Before  discussing numerical results, we  prove the convergence of the basic algorithms presented in this section.

\section{Theoretical performance}

 In this section, we establish the bound for the error of approximating the solution of (\ref{L0}) with the iterates generated by the OT, OTP, ROT or ROTP  under the  restricted isometry property (RIP). In compressed sensing language, we prove the success of signal recovery via these algorithms under the RIP. Our analysis allows the measurements of the signal to be inaccurate, and   we will point out at the end of this section  that our analysis is also valid when the target signal $x^*$ is not  precisely $k$-sparse. In particular, if the measurements   are accurate and the target signal   is $k$-sparse, our results claim  that the sequences generated by OT, OTP, ROT or ROTP converge to the target signal under the RIP. Let us first recall
 the  restricted isometry constant $\delta_K$ introduced by Cand\`es and Tao \cite{CT05}.

\begin{Def}\label{Def04}\emph{\cite{CT05, C08}} Given a matrix $A \in \mathbb{R}^{m\times n}$ with $m <n,$  the $K$th restricted isometry constant, denoted by  $\delta_K, $ is the smallest number $\delta \geq 0$ such that
$$ (1-\delta ) \|x\|^2_2 \leq \|Ax\|^2_2
\leq (1+\delta )\|x\|^2_2 $$ holds for all $K$-sparse vector
$x\in \mathbb{R}^n. $
\end{Def}

The following properties will be frequently used in later analysis.

\begin{Lem}
 \label{Lem-04-01}\emph{\cite{CT05, NT09, F11}}    Given  $u \in \mathbb{R}^n $ and the set $S \subseteq \{1,2, \dots, n\}$, one has
\begin{itemize}
\item[\textrm{(i)}]  $\| ((I-A^TA)v)_S\|_2  \leq \delta_t \|u\|_2  ~ \textrm{ if }|S\cup \emph{supp} (v) | \leq t. $

 \item[\textrm{(ii)}]  $ \|(A^Tu)_S\|_2 \leq \sqrt{1+\delta_t } \|u\|_2  ~ \textrm{ if } |S| \leq t. $
 \end{itemize}
\end{Lem}

\subsection{Analysis of OT and OTP algorithms} \label{subsection-04-01}

We first analyze the theoretical performance of the OT and OTP which provide a basic framework for the development of the  ROT and  ROTP and their variants. The main result for OT and OTP is summarized as follows.

\begin{Thm} \label{Thm-OIHT-01}  For every $k$-sparse vector $x$ satisfying $y= Ax+ \nu, $  if the restricted isometry constant of the matrix $A$ satisfies  $\delta_{2k} <  \tau^* \approx 0.5349,  $  where $\tau^* $ is the real root of the univariate equation $ \tau^3+ \tau^2 + \tau =1, $    then the iterates $\{x^p\}  $ generated by OT or OTP  approximate $x$ with $\ell_2$-error \begin{equation}  \label{err-111}   \|x^{p}-x \|_2 \leq  \rho^p \| x^0-x\|_2   +   C \|\nu\|_2,   \end{equation}
where $\rho$ and $C$ are constants given by $$ \rho  =  \delta_{2k}  \sqrt{ \frac{1+\delta_{2k}} {1-\delta_{2k} } }  < 1 , ~~  C=\frac{3+\delta_{2k}} {(1-\rho)\sqrt{1-\delta_{2k}}}.  $$ In particular, when $\nu=0,$ i.e., $y=Ax$, the iterates $\{x^p\}  $ generated by OT or OTP  converge to $x.$
\end{Thm}

\emph{Proof.}   Let $x^p$ be the current iterate, generated by OT or OTP, which is $k$-sparse. Denote by   $ u^p = x^p+ A^T (y-A x^p) $ and $ {\cal W}^{(k)} =\{w:  ~ \textrm{\textbf{e}}^T w =k,~ w \in \{0,1\}^n\}. $ Note that $y= Ax+ \nu. $ So \begin{equation} \label{EE01}  x-u^p  = (I- A^T A) (x- x^p) -A^T \nu. \end{equation}

(I)  We first analyze the OT algorithm.  Note that $w^*(u^p)\in{\cal W}^{(k)} $  is a minimizer of the problem (\ref{HT-QP}). Thus
\begin{equation} \label{equation-d}     \|y-A[u^p \otimes  w^*(u^p)] \|_2 \leq  \|y-A(u^p \otimes  w ) \|_2  ~ \textrm{    for any } w \in {\cal W}^{(k)}.  \end{equation}  By the structure of the OT algorithm,  $ x^{p+1} =  u^p\otimes  w^*(u^p)   $  and thus  $  x^{p+1}$ is a $k$-sparse vector with  $ \textrm{supp} (x^{p+1}) \subseteq \textrm{supp} (w^*(u^p)) . $  Since $x$ is a $k$-sparse vector, there exists a $k$-sparse binary vector $\widehat{w} \in {\cal W}^{(k)}$ such that $ \textrm{supp}(x) \subseteq  \textrm{supp} (\widehat{w}), $ and hence   $x \otimes (\textrm{\textbf{e}}- \widehat{w}) =0.$  Then it follows from (\ref{equation-d}) that
\begin{equation} \label{EE00} \|y-A x^{p+1}\|_2  \leq  \|y-A(u^p \otimes  \widehat{w} ) \|_2. \end{equation}
Note that $x^{p+1}-x$ is a $(2k)$-sparse vector. By Lemma \ref{Lem-04-01}, we have $\|A(x-x^{p+1})\| \geq \sqrt{1-\delta_{2k}} \|x-x^{p+1}\|. $   Thus   $$   \|y-A x^{p+1}\|_2   =  \|A(x- x^{p+1}) +\nu \|_2 \geq  \sqrt{1-\delta_{2k}} \| x-x^{p+1}\|_2 - \|\nu \|_2. $$ Merging this inequality with (\ref{EE00}) leads to
 \begin{equation} \label{EE02}  \|x^{p+1}-x\|_2  \leq    \frac{1}{ \sqrt{ 1-\delta_{2k}  }   }  (\|y-A(u^p \otimes \widehat{w} ) \|_2 + \|\nu\|_2).  \end{equation}
We now estimate the right-hand side of (\ref{EE02}).
By the choice of $\widehat{w}  $ and noting that   $ \textrm{supp} (x)  \subseteq \textrm{supp} (\widehat{w}), $ we see that $|\textrm{supp} (\widehat{w}) \cup \textrm{supp} (x-x^p)|\leq 2k.$ Therefore, by (\ref{EE01}) and Lemma \ref{Lem-04-01}, we have
\begin{eqnarray}   \label{estimate-01}  \|  (x- u^p) \otimes  \widehat{w}  \|_2  & = & \left\| [(I-A^TA) (x-x^p)] \otimes  \widehat{w}  - (A^T \nu ) \otimes  \widehat{w} \right \|_2 \nonumber \\  &  \leq   &  \left\| [(I-A^TA) (x-x^p)]    _{\textrm{supp}(\widehat{w})}   \right \|_2  +  \|(A^T \nu )_{\textrm{supp}( \widehat{w})} \|_2 \nonumber \\
 & \leq &   \delta_{2k} \|x-x^p\|_2  + \sqrt{1+\delta_k} \| \nu \|_2.
 \end{eqnarray}
As $(x- u^p) \otimes  \widehat{w} $ is a $k$-sparse vector, we obtain
\begin{eqnarray}   \|y-A(u^p \otimes  \widehat{w}) \|_2   & = &  \| \nu + A(x- u^p \otimes  \widehat{w} )\|_2   \nonumber \\
 & = &  \| \nu + A [(x-u^p)\otimes  \widehat{w}  + x  \otimes (\textrm{\textbf{e}}- \widehat{w}) ]\|_2  \nonumber\\
&   =  &   \| \nu+ A [ (x-u^p)\otimes  \widehat{w} ]    \|_2   \nonumber\\
&   \leq   &  \|\nu \|_2 +  \sqrt{1+\delta_k}   \|(x-u^p) \otimes  \widehat{w} \|_2  \nonumber\\
& \leq  &  \delta_{2k} \sqrt{1+\delta_k} \|x-x^p\|_2 +  (2+\delta_k) \|\nu \|_2.   \label{EE03}
\end{eqnarray}
The third equality above follows from the fact $x  \otimes (\textrm{\textbf{e}}- \widehat{w}) =0. $ The first inequality above  follows from Definition \ref{Def04} with the fact $(x-u^p) \otimes  \widehat{w} $ being  $k$-sparse. The last inequality follows from (\ref{estimate-01}).
Note that $\delta_k \leq \delta_{2k}.$ Combining (\ref{EE02}) and  (\ref{EE03}) yields
\begin{equation} \label{ERR}
\|x^{p+1}-x\|_2    \leq      \delta_{2k}  \sqrt{ \frac{1+\delta_{k}} {1-\delta_{2k} } }    \|x-x^p\|_2   + \frac{3+\delta_k}{\sqrt{1-\delta_{2k} } } \|\nu \|_2 \leq  \rho  \|x-x^p\|_2  + \frac{3+\delta_{2k}}{\sqrt{1-\delta_{2k} } } \|\nu \|_2,
\end{equation}
where  $   \rho  :=     \delta_{2k}  \sqrt{ \frac{1+\delta_{2k}} {1-\delta_{2k} } }  < 1     $ which  is ensured by $\delta_{2k} < \tau^*,$ where $\tau^* ~(\approx 0.5349)  $ is the positive real root of the univariate equation $ \tau^3+ \tau^2 + \tau =1. $ The result (\ref{err-111}) follows immediately from (\ref{ERR}) and the fact $\sum_{i=1}^\infty \rho^i =\frac{1}{1-\rho}.$

 (II) We now consider the  OTP algorithm, which generates the next iterate $x^{p+1}$ by performing the  orthogonal project step   $$ \min_{z} \{\|y-A z \|_2^2:   ~ \textrm{supp} (z) \subseteq  \textrm{supp} (u^p \otimes w^*(u^p))   \}, $$   which implies that
$$  \| y- A x^{p+1}\|_2      \leq   \|y-A(u^p \otimes  w^*(u^p)  ) \|_2  \leq    \|y-A(u^p \otimes  \hat{w }) \|_2
$$
 where the last inequality follows from  (\ref{equation-d}) by setting $w = \widehat{w}. $
Therefore, the iterate $x^{p+1}$ generated by the OTP also satisfies the relation (\ref{EE00}). Repeating the same proof above for the OT algorithm,
 we see that (\ref{ERR}) remains valid for the OTP with the same constant $\rho <1. $

 In particular, when $\nu=0,$ it follows immediately from (\ref{err-111}) that the sequence $\{x^p\}$ generated by OT or OTP converges to $x.$

 \vskip 0.08in

The  result above is shown under the condition $ \delta_{2k} < \tau^*. $ The RIP condition has been widely used in the theoretical analysis of various thresholding algorithms.  For instance,  the convergence of the HTP  was shown under the condition $\delta_{3k}<1/\sqrt{3}$ (see \cite{F11, FR13}),  and that of the IHT algorithm with a stepsize taken in  $  (\frac{1}{2(1-\delta_{2k})}, \frac{1}{1+\delta_{2k}}) $ was shown under the condition $\delta_{2k} < 1/3$  (see \cite{GK09, BD10, B12, FR13}).

In the case $\nu =0, $  the convergence rate of $\{x^p\}$ in Theorem \ref{Thm-OIHT-01} can be further enhanced, as shown by the next corollary.

\begin{Cor} \emph{(Local convergence rate)} For every $k$-sparse vector $x$ with $y=Ax,$ under the same condition of Theorem \ref{Thm-OIHT-01}, there exists an integer number $ \widehat{p} $ such that for all $p \geq  \widehat{p}, $
$$ \|x^{p}-x\|_2  \leq  ( \rho ^*)^p    \|x^0-x\|_2 ,  $$
 where \begin{equation} \label{en-bound}  \rho^*  : = \delta_{k}  \sqrt{ \frac{1+\delta_{k}} {1-\delta_{k} } }   \leq  \rho=  \delta_{2k}  \sqrt{ \frac{1+\delta_{2k}} {1-\delta_{2k} } } <1 . \end{equation}
\end{Cor}

\emph{Proof.}    By Theorem \ref{Thm-OIHT-01}, when $\nu =0,$ the sequence $\{x^p\} $  generated by OT or OTP algorithm converges to $x.$  Thus there is a sufficiently large integer number $ \widehat{p} $ such that $ \textrm{supp} (x) \subseteq \textrm{supp} (x^{p}) \textrm{  for any } p \geq  \widehat{p}.  $   In fact, if there is an index $i_0\in \textrm{supp} (x)$ such that $ i_0\notin \textrm{supp} ( x^p), $   then $ \|x-x^p \|_2 \geq |x_{i_0}|,$ contradicting to the fact $ x^p\to x $ as $p \to \infty . $      Therefore, $x- x^{p+1}$ and $ x-x^p $ must be  $k$-sparse for all $p \geq \hat{p}.$  The left-hand side of (\ref{equation-d}) larger than or equal to $\sqrt{1-\delta_k} \|x-x^{p+1}\|_2. $
For  $p \geq  \widehat{p}$, picking a vector in ${\cal W}^{(k)}, $ denoted by $\widehat{w}^p, $ which satisfies that  $ \textrm{supp}( x^p) \subseteq \textrm{supp} (\widehat{w}^p). $ This implies that
  $ \textrm{supp} (x)  \subseteq \textrm{supp} (\widehat{w}^p)  $ for all $p \geq \widehat{p}. $  Therefore, $x \otimes (\textrm{\textbf{e}}- \widehat{w}^p) =0$ for  all $p \geq \widehat{p}. $ Replacing the vector $ \widehat{w} $ in the proof of Theorem \ref{Thm-OIHT-01} with $\widehat{w}^p, $    the inequality (\ref{estimate-01}) can be improved to $  \|  (x- u^p) \otimes  \widehat{w}^p  \|_2   \leq  \delta_{k} \|x-x^p\|_2$ due to the fact $\nu =0 $ and  $ |\textrm{supp} (\widehat{w}^p) \cup \textrm{supp} (x- x^p)|\leq k. $  The estimation (\ref{EE03}) can be improved to $$   \|y-A(u^p \otimes \widehat{w}^p  ) \|_2
  \leq  \delta_{k}   \sqrt{1+\delta_k}   \|x-x^p\|_2.
$$ Therefore, from the proof of   Theorem \ref{Thm-OIHT-01},  we have
$$  \|x^{p+1}-x\|_2  \leq   \delta_{k}  \sqrt{\frac{1+\delta_{k}} {1-\delta_{k} } }   \|x-x^p\|_2 , ~  p\geq \widehat{p} . $$
 Since $\delta_k \leq \delta_{2k}< \tau^*,$  we immediately see the relation in (\ref{en-bound}).

This result indicates that the local convergence speed of the OT and OTP may  actually be faster than what Theorem \ref{Thm-OIHT-01} claims.

\subsection{Analysis of ROT and ROTP algorithms}

We now  analyze the ROT and ROTP algorithms which are the tightest convex relaxation counterparts  of the OT and OTP, respectively. Note that the solution $w^p$ of the relaxation problem in (\ref{HT-QP-Relax})  may not be exactly binary (and hence may not be $k$-sparse). So the analysis in Section \ref{subsection-04-01}, based on the optimal $k$-thresholding indicator $w^*(u^p),$ does not apply to the ROT and ROTP for which a non-trivial   analysis will be provided in this section.  We first give a few useful lemmas.

\begin{Lem} \label{Lem4422} Let  $z\in \mathbb{R}^n $  be a given vector.  Then for any $\hat{z} \in {\cal H}_k(z), $   one has
$$ \|z- \hat{z} \|_2^2 \leq \|z-x\|_2^2-\|(z-x)_S\|^2_2 $$ for any $k$-sparse vector $x\in \mathbb{R}^n$ with $S= \emph{supp} (x).  $

\end{Lem}

\emph{Proof.}  Since ${\cal H}_k(z)$ retains the largest $k$ magnitudes of $z,$  for any $\hat{z} \in {\cal H}_k (z), $    $ \|z-\hat{z}\|_2^2$ is the sum of the squares of   the $n-k$ smallest magnitudes of $z,$ which  must be smaller than or equal to the sum of the squares of any $n-k$ components of $z. $ So   $ \|z- \hat{z} \|_2^2 \leq  \| z_{\overline{S}}\|_2^2  $   for any set $ S \subseteq \{1, \dots, n\} $ with $|S| \leq k , $ where  $\overline{S} = \{1, \dots, n\}\backslash S.$
Let $x$ be any $k$-sparse vector with $S=\textrm{supp}(x). $ As $x_{\overline{S}}=0 $ and $|S| \leq k,$  by setting $S =\textrm{supp} (x) $ in the inequality above,  we immediately have
  $$ \|z-\hat{z}\|_2^2 \leq \| z_{\overline{S}}\|_2^2  = \| (z-x)_{\overline{S}}\|_2^2 =  \|z-x\|_2^2-\|(z-x)_S\|^2_2,   $$   as desired. ~~ $\Box $

\begin{Lem} \label{Lem4433}  Let $u^p \in \mathbb{R}^n $ be a given vector, and let  $x\in \mathbb{R}^n $ be a  $k$-sparse vector with   $S= \emph{supp} (x). $  Let $w^p $ be a solution to the  problem (\ref{HT-QP-Relax}). Then for  any vector $v\in{\cal H}_k(u^p\otimes w^p) $ with $ S^{p+1} = \emph{supp} (v), $ one has
$$\|x- v \|_2 \leq \| (u^p\otimes w^p- x)_{S^{p+1} \cup S}\|_2 + \|(u^p\otimes w^p-x)_{S^{p+1}\backslash S}\|_2 . $$
\end{Lem}

   \emph{Proof. }Let $x \in \mathbb{R}^n$ be a  $k$-sparse vector with $S=\textrm{supp} (x). $ By setting $z = u^p\otimes w^p   $ in Lemma 4.5, for any $v\in{\cal H}_k(u^p\otimes w^p), $ we have
$$ \| u^p\otimes w^p  - v\|_2^2 \leq \| u^p\otimes w^p -x\|_2^2-\|( u^p\otimes w^p -x)_S\|_2^2. $$
The left-hand side can be written as
$$  \|u^p \otimes w^p -  v \|^2_2
 = \|u^p \otimes w^p  - x   \|^2_2  +  \|  x -v \|^2_2   +2 (x-v)^T ( u^p \otimes w^p -x ) .
  $$
  Note  that $\textrm{supp}\left( x- v \right) \subseteq  \textrm{supp} (v) \cup\textrm{supp} (x)=  S^{p+1}\cup S,$  where $ S^{p+1} = \textrm{supp} (v). $  Combining the  two relations above yields
 \begin{eqnarray}  \label{Est-01}
 \|  x -  v \|^2_2
  & \leq &    - \|(u^p\otimes w^p -x )_S\|_2^2 -   2 (x- v )^T ( u^p \otimes w^p -x ) .  \nonumber \\
                         & =  &    - \|(u^p\otimes w^p -x )_S\|_2^2 - 2  [(x- v )_{ S^{p+1}\cup S }]^T  [ u^p \otimes w^p -x ] _{ S^{p+1}\cup S } \nonumber \\
                         & \leq   &    - \|(u^p\otimes w^p-x )_S\|_2^2+ 2\|x- v \|_2 \|[ u^p \otimes w^p -x ]_{ S^{p+1}\cup S }\|_2.
  \end{eqnarray}
Note that the positive root of the quadratic function (in variable t)
$$ t^2   -  2 t \|( u^p \otimes w^p -x )_{  S^{p+1}\cup S}\|_2 + \|(u^p\otimes w^p -x )_S\|_2^2  =0$$
is given as follows:
\begin{eqnarray*}  t^*  & =  & \frac{2 \| (u^p \otimes w^p -x ) _{ S^{p+1}\cup S }\|_2+ \sqrt{4 \|( u^p \otimes w^p -x ) _{S^{p+1}\cup S  }\|_2^2-4\|(u^p\otimes w^p-x )_S\|_2^2} }{2} \\
 & = & \| (u^p\otimes w^p- x)_{S^{p+1}\cup S }\|_2 +  \|(u^p\otimes w^p-x)_{S^{p+1}\backslash S}\|_2
\end{eqnarray*}
The inequality (\ref{Est-01})  implies that  $ \|  x - v\|_2 \leq t^*,   $  as desired.

\vskip 0.07in

The next lemma has been shown in the proof of Theorem \ref{Thm-OIHT-01}. See   (\ref{EE03}) for details.

\begin{Lem} \label{Lem4444} Let $x\in \mathbb{R}^n$ be a  $k$-sparse vector satisfying $y= Ax +\nu. $ Let $x^p \in \mathbb{R}^n  $ and  $u^p = x^p + A^T (y-Ax^p) . $   Then for any $ \widehat{w} \in {\cal W}^{(k)} $ satisfying $\textrm{supp} (x) \subseteq  \widehat{w}, $   one has
$$ \|y-  A(u^p\otimes \widehat{w})\|_2  \leq  \delta_{2k}  \sqrt{1+\delta_k} \| x- x^p\|_2 + (2+\delta_k) \|\nu \|_2.  $$

\end{Lem}

We now prove the main result for ROT and ROTP algorithms.

\begin{Thm} \label{Thm-OIHT-02}
  Let $x$ be a $k$-sparse vector satisfying $y = Ax+ \nu. $ Suppose that the restricted isometry constant of the matrix $A$ satisfies  $\delta_{3k} \leq   1/5.$  Then the iterates $\{x^p\}, $ generated by ROT or ROTP, approximate  $x$ with error
   \begin{equation} \label{error-a}  \|x^p-x \|_2 \leq  \varrho^p  \| x^0-x\|_2 + C^* \|\nu\|_2,  \end{equation}
where, for ROT, the constants $\varrho$ and $C^*$ are given as  $$ \varrho : = ( \delta_{2k} +  2  \delta_{3k} )\sqrt{ \frac{ 1+ \delta_k}  {  1-\delta_{2k} } } + \delta_{3k}    < 1 ,  ~~ C^* =\frac{1}{1-\varrho} \left(\frac{ 5+3\delta_k}{\sqrt{1-\delta_{2k}}}+\sqrt{1+\delta_k}\right),  $$   and for ROTP the constants $\varrho$ and $C^*$ are given as
  \begin{equation}\label{CCC-aa} \varrho  =\frac{1}{\sqrt{1-\delta_{2k}^2}}\left( ( \delta_{2k} +  2  \delta_{3k} )\sqrt{ \frac{ 1+ \delta_k}  {  1-\delta_{2k} } } + \delta_{3k} \right)    < 1 , \end{equation}
\begin{equation} \label{CCC-bb}  C^*=  \frac{1}{1-\varrho} \left(\frac{5+3\delta_k}{(1-\delta_{2k})\sqrt{1+\delta_{2k}}}+  \frac{\sqrt{1+\delta_k}} {  \sqrt{ 1-\delta_{2k}^2 } } + \frac{  \sqrt{1+\delta_k} }{ 1- \delta_{2k} }\right).  \end{equation}
In particular, when $\nu=0$  (i.e., $y=Ax$), the sequence $\{x^p\}$ generated by ROT or ROTP converges to $x.$
\end{Thm}

 \emph{Proof.}   (I) We first analyze the ROT.   At   $x^p$, the ROT  generates the vector $w^p$ by solving the optimization  problem  (\ref{HT-QP-Relax}) with   $u^p= x^p+ A^T (y-Ax^p). $   Then the next iterate is given by
$ x^{p+1} \in  {\cal H}_k(u^p \otimes w^p). $
Denote by  $ S^{p+1} =\textrm{supp} (x^{p+1}) .$
Since  $x$ is a $k$-sparse vector with  $S=\textrm{supp} (x),$   by Lemma \ref{Lem4433}, we have
 \begin{equation} \label{Est-111} \|  x -x^{p+1} \|_2 \leq  \left\|( u^p \otimes w^p -x ) _{  S^{p+1}\cup S}\right\|_2 + \left\|( u^p \otimes w^p -x ) _{ S^{p+1} \backslash S}\right\|_2  .  \end{equation}
We now estimate the upper bound for the  right-hand side of the above inequality.  By using (\ref{EE01}) and noting that $ x_{ S^{p+1} \backslash S} =0   $ and $ 0\leq w^p\leq \textrm{\textbf{e}},$ we have
\begin{eqnarray}  \label{Bound-01} \left \|( u^p \otimes w^p -x ) _{ S^{p+1} \backslash S} \right\|_2 &  = &  \left \|( u^p \otimes w^p ) _{ S^{p+1} \backslash S} \right\|_2   =  \|[( u^p-x) \otimes w^p ] _{ S^{p+1}\backslash S}\|_2  \nonumber \\
                         &  = & \left \| [(A^T \nu - (I-A^TA) (x- x^p ))\otimes w^p]_{ S^{p+1}\backslash S} \right\|_2  \nonumber \\
     & \leq   & \left \| [(I-A^TA) (x-x^p )]_{ S^{p+1} \backslash S}\right\|_2  + \| (A^T \nu)_{S^{p+1} \backslash S}\|_2 \nonumber \\
     & \leq  &  \delta_{3k} \|x^p-x\|_2 +\sqrt{1+\delta_k} \|\nu\|_2,
\end{eqnarray}
where  the last inequality follows from Lemma \ref{Lem-04-01} due to the fact $|\textrm{supp} ( x-x^p ) \cup (S^{p+1}\backslash S) |\leq 3k $ and $|S^{p+1} \backslash S|\leq k.$
Using $y=Ax+\nu, $ we  have
 \begin{eqnarray*}   & &   \|y-A(u^p \otimes  w^p   ) \|_2  \\
  & & =     \|A( u^p \otimes  w^p  -x) -\nu  \|_2    \\
                             & & =  \left\|A [  (u^p \otimes  w^p-x )_{ S^{p+1}\cup S }] + A [ ( u^p \otimes  w^p -x )_{\overline{ S^{p+1}\cup S}} ] - \nu  \right \|_2   \\
         &  &  \geq   \|A [  (u^p \otimes  w^p -x )_{ S^{p+1} \cup  S} ] \|_2- \left\| A [  (u^p \otimes  w^p-x )_{\overline{ S^{p+1}\cup S}} ] \right\|_2  -\|\nu\|_2 \\
            &   & \geq     \sqrt{1-\delta_{2k}}  \|  (u^p \otimes  w^p-x )_{ S^{p+1} \cup S  } \|_2 -  \left\| A [  (u^p \otimes  w^p -x )_{\overline{  S^{p+1}\cup  S }} ]\right \|_2 -\|\nu\|_2,
\end{eqnarray*}
and thus
\begin{equation}  \label{Est-02}    \left\| (u^p \otimes  w^p -x )_{ S^{p+1} \cup  S } \right \|_2
  \leq \frac{1} { \sqrt{1-\delta_{2k}}  } \left(  \| y- A(u^p \otimes w^p) \|_2   +  \mathcal{T}   + \|\nu\|_2\right) , \end{equation}
  where
  $$  \mathcal{T} :=  \| A [ ( u^p \otimes  w^p -x )_{\overline{S^{p+1}\cup S}}  ] \|_2 . $$
  Let $\widehat{w} \in {\cal W}^{(k)} $ be a vector such that $S =\textrm {supp} (x) \subseteq \textrm{supp} (\widehat{w}),$ which implies that $x \otimes (\textrm{\textbf{e}}-\widehat{w})  =0 .  $  Since $w^p$ is an optimal solution to (\ref{HT-QP-Relax}),   we have
\begin{equation} \label{basic}  \| y-A(u^p \otimes w^p)\|_2 \leq \| y- A (u^p \otimes \widehat{w}) \|_2 \leq \delta_{2k}\sqrt{1+ \delta_k} \|x^p-x\|_2 +(2+\delta_k) \|\nu\|_2,  \end{equation}  where the last inequality follows from Lemma \ref{Lem4444}.
Combining (\ref{Est-111}), (\ref{Bound-01}) , (\ref{Est-02}) and (\ref{basic}), we have
\begin{eqnarray} \label{Est-03}  \|  x -x^{p+1} \|_2  & \leq  &  \frac{1} { \sqrt{1-\delta_{2k}}  } \left[  \delta_{2k}\sqrt{1+ \delta_k} \|x^p-x\|_2    + \mathcal{T}  \right]   + \left[ \frac{3+\delta_k} {\sqrt{1-\delta_{2k}}}+ \sqrt{1+\delta_k}\right] \|\nu\|_2 \nonumber  \\
 & &   +  \delta_{3k} \|x^p-x\|_2.
 \end{eqnarray}
In the remainder of the proof, we estimate the term  $\mathcal{T} . $
  Since  $ x_{\overline{S\cup S^{p+1}}} =0, $  $ \mathcal{T} $ can be written as
$$ \mathcal{T}  = \| A [( u^p \otimes  w^p) _{\overline{ S^{p+1}\cup S}}  ]  \|_2   = \| A [(u^p-x) \otimes  w^p]_{\overline{  S^{p+1}\cup S}}  \|_2 . $$
   Let $ \left |\overline{ S^{p+1}\cup  S} \right |  =  (\widehat{n}-1)  k + \ell, $ where $\widehat{n} $ and $  \ell $ are integer numbers and $0\leq \ell < k. $
    Let $$\overline{ S^{p+1} \cup S} = S_1 \cup S_2 \cup  \cdots \cup S_{\widehat{n} -1 } \cup S_{\widehat{n} } $$ be the disjoined partition of $ \overline{ S^{p+1}\cup S }, $ satisfying the following properties:
  \begin{itemize}

  \item [(i)]   $ S_i \cap S_j = \emptyset $ for $ i \not= j, $  and  $  |S_i|=k \textrm{ for all } i =1, \dots, \widehat{n}-1  $ and $ |S_{\widehat{n}}|= \ell< k.$

   \item[(ii)]   $   {S_1}$ is the index set for  the $k$ largest elements  in  the set $ \{(w^p)_i: i \in \overline{ S^{p+1}\cup S } \},$   $S_2$ is the index set for the second $k$ largest elements in  this set, and so on.
       \end{itemize}
       Thus the vector $(w^p)_{\overline{  S^{p+1}\cup S}} $ is decomposed as  $$ (w^p)_{\overline{S^{p+1}\cup S}} = (w^p)_{S_1} +  \cdots  + (w^p)_{S_{\widehat{n}-1}}+  (w^p)_{S_{\widehat{n}} }.  $$
       Sorting  the components of $w^p$ supported on $S_i$ $ (i=1, \dots, \hat{n}-1) $ into descending order, and denote such  ordered components by $\alpha^{(i)}_1 \geq \alpha^{(i)}_2 \geq \cdots \geq \alpha^{(i)}_k , $ and denote the ordered components of $w^p$ supported on $S_{\hat{n}}$ by  $\alpha^{(\hat{n})}_1 \geq \alpha^{(\hat{n})}_2 \geq \cdots \geq \alpha^{(\hat{n})}_\ell . $ Thus  $\alpha^{(i)}_1$ denotes the largest entries of $w^p$ on the support $S_i $ for $i=1,\dots, \hat{n}, $    $ \alpha_{k} ^{(i)} $ denotes the smallest entry of $w^p$ on the support $S_i$ for $i=1, \dots, \hat{n}-1,$ and $ \alpha^{(\hat{n})}_{\ell}$ denotes the smallest component of $w^p$ supported on $ S_{\hat{n}}. $   By this notation,  sorting the components of the vector $ (w^p)_{\overline{S^{p+1}\cup S}} $ supported on $ \overline{S^{p+1}\cup S}$  into descending order, we obtain the  sequence as follows:
       $$  \overbrace{\alpha^{(1)}_1 \geq \alpha^{(1)}_2  \geq \cdots \geq \alpha^{(1)}_k  }  \geq   \overbrace{\alpha^{(2)}_1 \geq \alpha^{(2)}_2  \geq \cdots \geq \alpha^{(2)}_k  } \geq \cdots \geq  \overbrace{\alpha^{(\widehat{n})}_1 \geq \alpha^{(\widehat{n})}_2  \geq \cdots \geq \alpha^{(\widehat{n})}_\ell  }.   $$
We now  prove that
 \begin{equation}  \label{Delta}  \Delta :=  \sum _{ i=1}^{\widehat{n}}  \alpha^{(i)}_1  \leq 2-\frac{1}{k} < 2. \end{equation}
For each $i, $ the largest entry of  $ (w^p)_{S_{i+1} }  $  is smaller than or equal to the smallest entry of $ (w^p)_{S_i},   $ i.e., $ \alpha^{(i)}_k \geq \alpha^{(i+1)}_1 $ for $ i = 1, \dots, \widehat{n}-1. $ So we immediately see that
$$ \Delta    =    \alpha^{(1)}_1 +  \alpha^{(2)}_1  + \cdots + \alpha^{( \widehat{n} )}_1 \leq  \alpha^{(1)}_1 + \alpha^{(1)}_k  + \cdots + \alpha^{(\widehat{n}-1 )}_k  \leq 1+  \sum_{i=1}^ {\widehat{n}-1} \alpha^{(i)}_k,
$$     where the last inequality follows from $\alpha^{(1)}_1 \leq 1$ due to the fact $ 0\leq w^p \leq \textrm{\textbf{e}}. $  Note that $   \alpha^{(i)}_k \leq  \alpha^{(i)}_{k-1} \leq \cdots \leq  \alpha^{(i)}_2.  $  Thus $$ \sum_{i=1}^ {\widehat{n}-1} \alpha^{(i)}_k \leq \sum_{i=1}^ {\widehat{n}-1} \alpha^{(i)}_{k-1} \leq \cdots \leq \sum_{i=1}^ {\widehat{n}-1} \alpha^{(i)}_2.   $$ So it follows from the inequalities above  that
 $  \Delta   \leq  1+  \sum_{i=1}^ {\widehat{n}-1} \alpha^{(i)}_j $ for $ j= 2, \dots, k. $
Adding these $k-1 $ inequalities to the equality $\Delta =  \sum _{ i=1}^{\widehat{n}}  \alpha^{(i)}_1  $ yields
\begin{eqnarray*}  k \Delta  & \leq  &  k-1 + \sum_{i=1}^ {\widehat{n} } \alpha^{(i)}_1  +  \sum_{i=1}^ {\widehat{n}-1 } \alpha^{(i)}_2 + \cdots + \sum_{i=1}^ {\widehat{n}-1 } \alpha^{(i)}_k     \leq    k-1 +  \sum_{j\in \overline{ S^{p+1} \cup S } }  (w^p)_j    \\
 &   \leq   &  k-1 + \|w^p\|_1 \\
 & = & 2k-1,  \end{eqnarray*}
 where the last inequality follows from the fact $ \|w^p\|_1 = \textrm{\textbf{e}}^Tw^p =k.  $ Thus
 (\ref{Delta}) holds.
 Define the vector $ v^{(i)}: =    [(u^p-x) \otimes  w^p]_{S_i},$  then $$  [(u^p-x) \otimes  w^p]_{\overline{ S^{p+1}\cup S }} = v^{(1)}+ v^{(2)} + \cdots +  v^ {(\widehat{n})}  . $$
       So the vector
   $[ (u^p-x) \otimes  w^p]_{\overline{ S^{p+1} \cup S }}$  is decomposed into $k$-sparse vectors  $v^{(i)} \in \mathbb{R}^n, i= 1, \dots,  \widehat{n} . $
Therefore,
\begin{equation} \label{TTT}  \mathcal{T} =   \left\| A \sum _{ i=1}^ {\widehat{n}} v^{(i)} \right\|_2 \leq \sum _{ i=1}^ {\widehat{n}} \|A v^{(i)} \|_2 \leq \sqrt{1+\delta_{k}} \sum _{ i=1}^{\widehat{n}} \|v^{(i)} \|_2,  \end{equation}
where the last inequality follows from Definition \ref{Def04} and the fact that  every $v^{(i)}$ is $k$-sparse. We now estimate the term    $\sum _{ i=1}  ^{\widehat{n}} \|v^{(i)} \|_2  .$
Note that
\begin{eqnarray*}   \|v^{(i)} \|_2   & =  &  \|[ (u^p-x) \otimes w^p]_{S_i}\|_2 \\
&  = &  \|[ (A^T \nu) \otimes w^p - ((I-A^TA)(x-x^p)) \otimes w^p]_{S_i} \|_2 \\
 &   \leq  & \|[(I-A^TA)(x-x^p)]_{S_i}  \otimes (w^p)_{S_i} \|_2  +  \|(A^T \nu) _{S_i} \otimes (w^p)_{S_i}\|_2 \\
  & \leq  &   \left( \max_{i\in S_i} (w^p)_i \right)  \| [(I-A^TA)(x-x^p)]_ {S_i} \|_2 + \left( \max_{i\in S_i} (w^p)_i \right)  \|(A^T \nu)_{S_i}\|_2 \\
  & \leq & \alpha^{(i)}_1 \delta_{3k} \|x-x^p \|_2 + \alpha^{(i)}_1 \sqrt{1+\delta_k} \|\nu \|_2,
  \end{eqnarray*}  where the last inequality follows from the fact $\alpha^{(i)}_1$  being the largest entry of $(w^p)_{S_i}$  and Lemma \ref{Lem-04-01} with  $ |S_i \cup \textrm{supp} (x-  x^p)| \leq 3k. $
 Thus
$$   \sum _{ i=1}  ^{\widehat{n}}  \|v^{(i)} \|_2   \leq   \Delta  \delta_{3k} \|x-x^p \|_2 +\Delta  \sqrt{1+\delta_k} \|\nu \|_2  \leq 2 \delta_{3k} \|x-x^p\|_2  +2 \sqrt{1+\delta_k} \|\nu \|_2.   $$
Merging (\ref{TTT}) and the inequality above leads to
$$ \mathcal{T} \leq 2 \delta_{3k} \sqrt{1+\delta_k}  \|x-x^p \|_2+ 2 (1+\delta_k) \|\nu \|_2. $$
Combining   (\ref{Est-03}) and  the above bound of $\mathcal{T} $ yields
  \begin{equation} \label{Est-04}  \|  x -x^{p+1} \|_2
 \leq   \varrho \|x-x^p \|_2    + \left[\frac{5+3\delta_k}{\sqrt{1-\delta_{2k}}}+\sqrt{1+\delta_k}\right]\|\nu\|_2,
   \end{equation}
  where  \begin{equation} \label{Est-05} \varrho:=   ( \delta_{2k} +  2  \delta_{3k} ) \sqrt{  \frac{1+ \delta_k}    { 1-\delta_{2k}  } } + \delta_{3k}   < 1 \end{equation}    under the condition $ \delta_{3k} \leq 1/5.$
In fact,   since $\delta_k \leq \delta_{2k} \leq \delta_{3k},$  we see that $    \varrho  \leq   3 \delta_{3k}  \sqrt{ \frac{ 1+\delta_{3k} } { 1-\delta_{3k}}  }  +  \delta_{3k}  <1  $ which is ensured by the condition $ \delta_{3k} \leq 1/5 . $    The bound (\ref{error-a})   immediately follows from (\ref{Est-04}) and (\ref{Est-05}).

(II) We now analyze the ROTP algorithm under the same assumption.  The ROTP solves the same optimization problem (\ref{HT-QP-Relax}) to obtain the vector $w^p.$  Let $ v $ be an arbitrary vector in ${\cal H}_k(u^p \otimes w^p). $ In ROT, $v$ is directly taken as the next iterate $x^{p+1} .$  The bound (\ref{Est-04}), which is shown for ROT,  holds for any vector $v$ in ${\cal H}_k(u^p\otimes w^p) .$ Therefore,
\begin{equation} \label{Est-06}  \|  x -v \|_2 \leq   \varrho  \|x-x^p \|_2 + C' \|\nu\|_2,
   \end{equation}
where $ \varrho$ is given  by (\ref{Est-05}) and $C'=\frac{5+3\delta_k}{\sqrt{1-\delta_{2k}}}+\sqrt{1+\delta_k} .$  The ROTP   uses $v$ as the intermediate point to compute the  iterate $x^{k+1}$ which is the  solution to the   orthogonal projection  problem
$$ \min_{z} \{ \| y- Az\|_2^2:  ~ \textrm{supp} (z) \subseteq S^{p+1}=  \textrm{supp} (v) \}. $$
Thus by optimality, the solution $x^{p+1}$ to this problem must satisfy that
$   [A^T (y-  A x^{p+1})]_{S^{p+1}} =0    $
which, by using $y=Ax+\nu, $  can be written as
$$  [(I-A^T A)(x-  x^{p+1})]_{S^{p+1}} =(x-  x^{p+1})_{S^{p+1}} + (A^T \nu)_{S^{p+1}} .   $$ This implies that
\begin{eqnarray*}      \|(x-  x^{p+1})_{S^{p+1}}\|_2
 &   \leq   &      \|[(I-A^T A)(x-  x^{p+1})]_{S^{p+1}}\|_2  +\|(A^T \nu)_{S^{p+1}}\|_2  \\
  &  \leq   &    \delta_{2k}  \| x-  x^{p+1}\|_2 +  \sqrt{1+\delta_k}  \|\nu\|_2.
\end{eqnarray*}
The   last equality follows from  Lemma \ref{Lem-04-01} due to the fact $ |\textrm{supp} (x-  x^{p+1})\cup S^{p+1} |\leq 2k  $ and $|S^{p+1}|\leq k.$
Noting that $ (x^{p+1})_{\overline{S^{p+1}}}=0$ and $ v_{\overline{S^{p+1}}}=0 ,$ we have
\begin{eqnarray*}  \|x-x^{p+1}\|_2^2 & = &  \|(x-x^{p+1})_{S^{p+1}} \|_2^2 + \|(x-x^{p+1})_{\overline{S^{p+1}}}\|_2^2\\
& = & \|(x-x^{p+1})_{S^{p+1}} \|_2^2  + \|(x-v)_{\overline{ S^{p+1}  }    }\|_2^2\\
& \leq & \delta_{2k}^2 \| x-  x^{p+1}\|_2^2+ 2 \delta_{2k} \sqrt{1+\delta_k}  \| x-  x^{p+1}\|_2  \| \nu\|_2 + (1+\delta_k) \|\nu\|_2^2\\
 & &   + \|(x-v)_{\overline{ S^{p+1}  }    }\|_2^2,
\end{eqnarray*}
and hence
$$ (1- \delta_{2k}^2) \|x-x^{p+1}\|_2^2 \leq 2 \delta_{2k} \sqrt{1+\delta_k}  \| x-  x^{p+1}\|_2  \| \nu\|_2 + (1+\delta_k) \|\nu\|_2^2
   + \|(x-v)_{\overline{ S^{p+1}  }    }\|_2^2 .  $$  This implies that
\begin{eqnarray*}    \|x-x^{p+1}\|_2
 &  \leq &  \frac{ 2\delta_{2k}\sqrt{1+\delta_k} \|\nu\|_2  + \sqrt{ 4 (1+\delta_k)\|\nu\|_2^2+   4 (1-\delta_{2k}^2)   \|(x-v)_{\overline{ S^{p+1}  } } \|_2^2  } } { 2 (1- \delta_{2k}^2)}\\
  &   \leq   & \frac{ 2\delta_{2k}\sqrt{1+\delta_k} \|\nu\|_2  +   2 \sqrt{1+\delta_k} \|\nu\|_2 +   2  \sqrt{1-\delta_{2k}^2}   \|(x-v)_{\overline{ S^{p+1}  } }  \|_2    } { 2 (1- \delta_{2k}^2) }\\
   &  \leq  &  \frac{  \sqrt{1+\delta_k} }{ 1- \delta_{2k} } \|\nu\|_2    +   \frac{1} { \sqrt{1-\delta_{2k}^2}  } \|(x-v)_{\overline{ S^{p+1}  } }  \|_2 \\
   &  \leq  &    \frac{  \sqrt{1+\delta_k} }{ 1- \delta_{2k} } \|\nu\|_2 + \frac{1}{\sqrt{1-\delta_{2k}^2}} \|x-v\|_2.
   \end{eqnarray*}
   Combining this inequality with (\ref{Est-06}) yields
  \begin{eqnarray*}  \|x-x^{p+1}\|_2  &   \leq &\frac{   \varrho\|x-x^p \|_2}{ \sqrt{ 1-\delta_{2k}^2 } }  + \left[\frac{  C'}{ \sqrt{ 1-\delta_{2k}^2 } } +\frac{  \sqrt{1+\delta_k} }{ 1- \delta_{2k} }\right] \|\nu\|_2=  \varrho' \|x-x^p \|_2 + C''\|\nu\|_2,
\end{eqnarray*}
 where $$C''= \frac{5+3\delta_k}{(1-\delta_{2k})\sqrt{1+\delta_{2k}}}+  \frac{\sqrt{1+\delta_k}} {  \sqrt{ 1-\delta_{2k}^2 } } + \frac{  \sqrt{1+\delta_k} }{ 1- \delta_{2k} } $$
and
  $$  \varrho'  :=\frac{1}{\sqrt{1-\delta_{2k}^2}}\left[ ( \delta_{2k} +  2  \delta_{3k} )\sqrt{ \frac{ 1+ \delta_k}  {  1-\delta_{2k} } } + \delta_{3k} \right] \leq \frac{ 3\delta_{3k}  } {  1-\delta_{3k} }  +\frac{ \delta_{3k}  }{ \sqrt{1-\delta_{3k}^2 } }   < 1, $$    where the first inequality follows from the  fact $\delta_{k} \leq \delta_{2k} \leq \delta_{3k} , $ and the last one follows from the condition  $\delta_{3k} \leq 1/5.$ Thus the error bound (\ref{error-a}), with constants (\ref{CCC-aa}) and (\ref{CCC-bb}), holds for ROTP.

 In particular, when $\nu =0$, i.e., $y=Ax$, the iterates $\{x^p\}$  generated by the ROT and ROTP converge to the sparse vector $x.$

\vskip 0.06in

\emph{Remark.}  In signal recovery scenarios,    the target signal $x$ is usually not exactly $k$-sparse and the measurements $y= Ax +\phi $  are also inaccurate, where $\phi$ is a noise vector. In such situations, we are interested in recovering the $k$ largest magnitudes of $x$ (which usually carry the most important information of the signal).  Our main results (Theorems \ref {Thm-OIHT-01} and \ref{Thm-OIHT-02}) can be immediately applied to such situations. In fact, let $S\subseteq \{1, \dots, n\}$ denote the index set for the $k$ largest magnitudes of the target signal $x.$ Note that  $$y=Ax + \phi = Ax_S + (Ax_{\overline{S}}+ \phi) = Ax_S+ \nu,  $$ where   $\nu = Ax_{\overline{S}}+ \phi  $ and $ \overline{S} = \{1, \dots, n \} \backslash S.$  The measurements $y$ of the original signal $x$ with noise $\phi$ can be seen as the measurements of the  $k$-sparse vector $x_S$ with noise $\nu = Ax_{\overline{S}}+ \phi .$ Therefore, Theorem  \ref {Thm-OIHT-01} claims that if $\delta_{2k}< \tau^* \approx 0.5349,  $  then the iterates $\{x^p\}  $ generated by OT or OTP  approximate $x_S$ with error \begin{equation}  \label{err-last}   \|x^{p}-x_S \|_2 \leq  \rho^p \| x^0-x_S\|_2   +   C \| Ax_{\overline{S}}+ \phi \|_2,   \end{equation}
where $\rho$ and $C$ are constants given in Theorem \ref{Thm-OIHT-01}. Also Theorem \ref{Thm-OIHT-02} shows that if $\delta_{3k}\leq 1/5,$ then the iterate $x^p$ generated by ROT or ROTP approximates $x_S$ with the error (\ref{err-last}), where the constants $ \rho$ and $  C $ are replaced  respectively  with $  \varrho $ and $  C^* $ that are given  in Theorem \ref{Thm-OIHT-02}.

\section{Numerical performance} \label{Section-NP}

Some preliminary experiments were performed to demonstrate the numerical behavior of the proposed algorithms. All matrices and sparse vectors are randomly generated.  The entries of matrices are assumed to be  i.i.d random variables which follow ${\cal N} (0,1),$ the standard normal distribution with zero mean and unit variance. The nonzero entries of the sparse vectors realized in our experiments are also assumed to follow such a distribution and the positions of nonzero entries are chosen randomly. All experiments were performed on a PC with the processor Intel(R) Core(TM) i5-3570 CPU @ 3.40 GHz and 8GB memory. All programs were written in MATLAB and the convex optimization problems were solved by using CVX developed by Grant and Boyd \cite{GB11} with solver `sedumi'.

\begin{figure} [htp]
$
\begin{array}{c}
\includegraphics [width=0.45\textwidth,
totalheight=0.225\textheight] {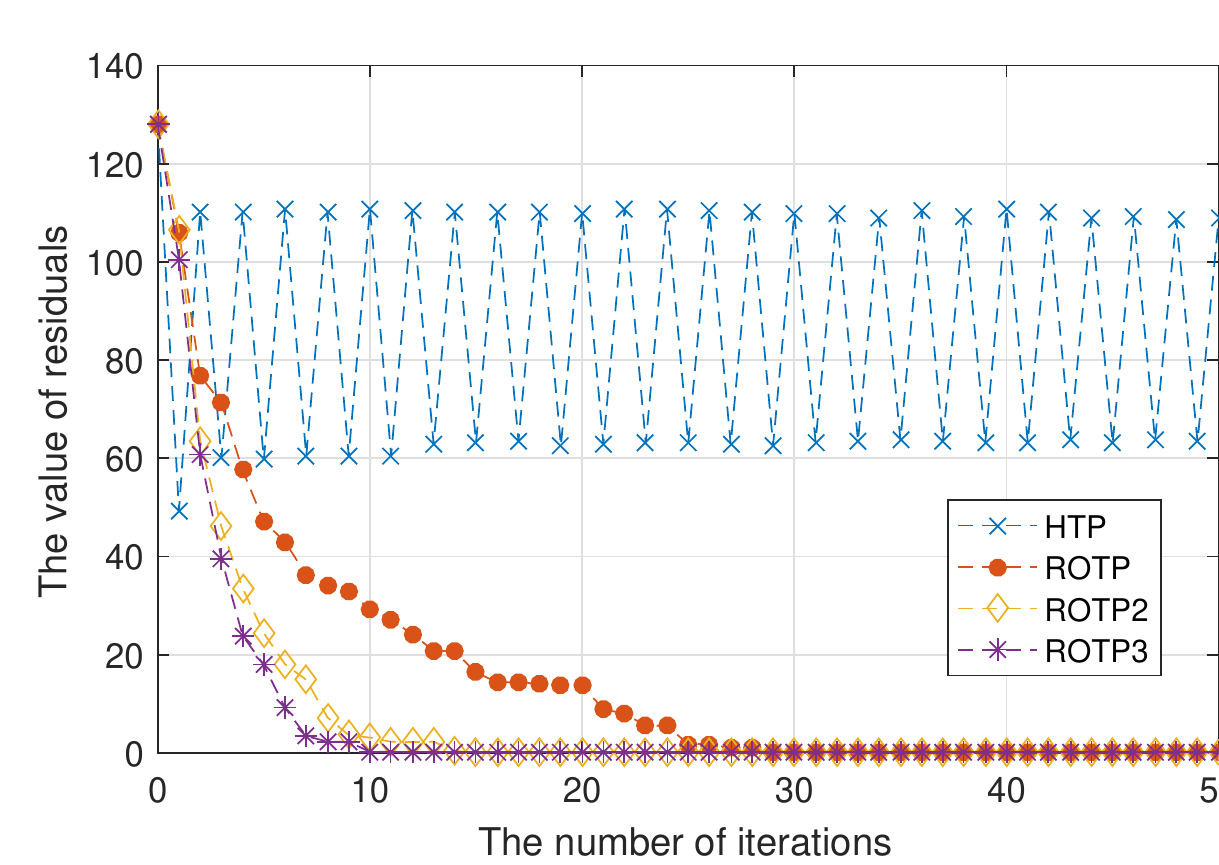}
\\
\textrm{(a)   Residual reduction}

\end{array}  $
\hfill
 $ \begin{array}{cc}
\includegraphics [width=0.45\textwidth,
totalheight=0.225\textheight] {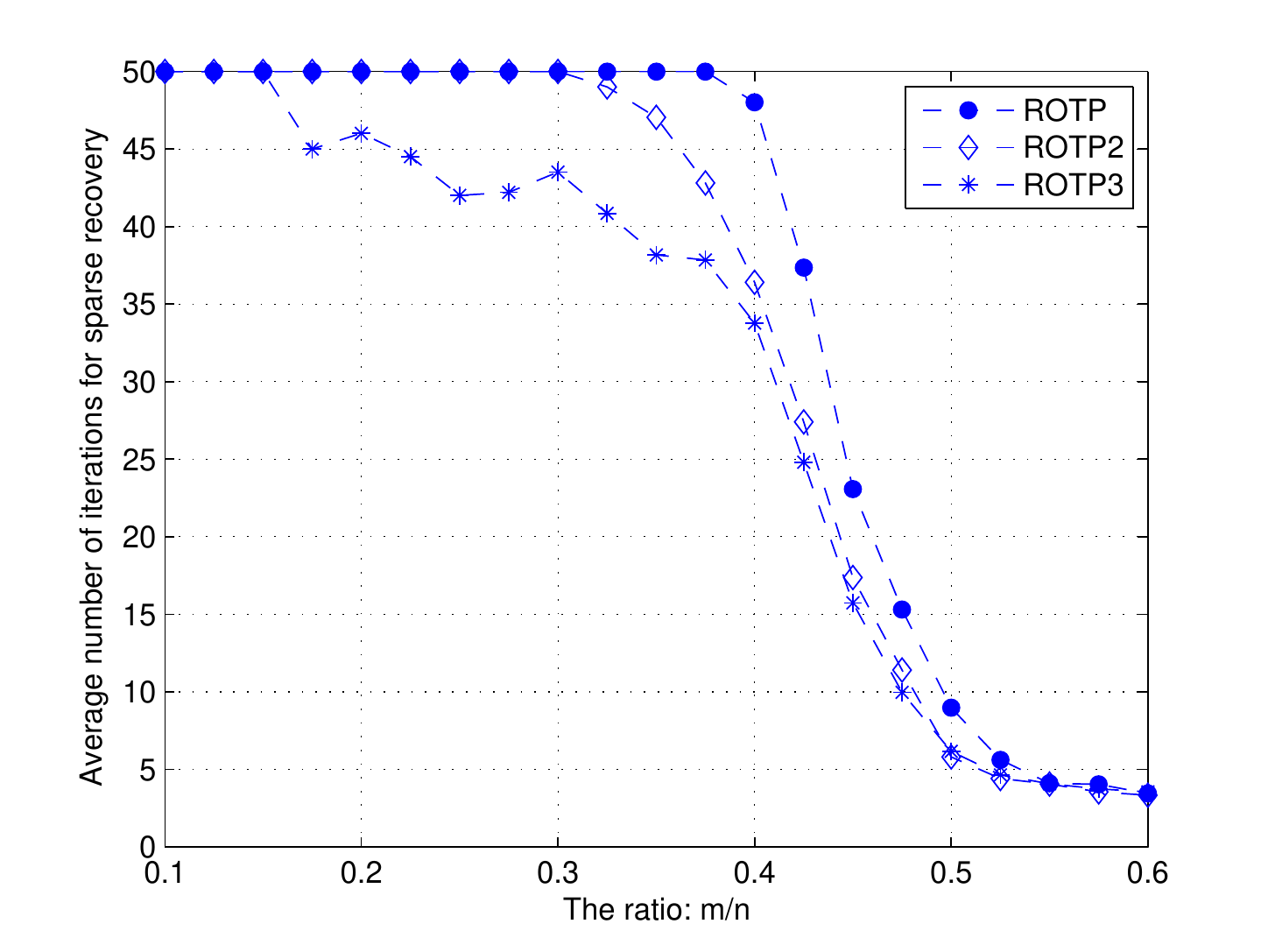} \\
 \textrm{(b)   Average no. of iter. for recovery}
\end{array}
$
 \caption{ Comparision of several algorithms in residual reduction, and the average number of iterations required for sparse recovery.  The maximum number of iterations is set as 50.}
\label{Fig-A}
\end{figure}

The first experiment was performed to illustrate the stableness of the proposed algorithms with respect to residual reduction.  We  generate a  random matrix $ A \in \mathbb{R}^{500 \times 1000}$  and a random sparse vector $x^*\in \mathbb{R}^{1000}$ with sparsity level $k= 120$ (i.e., $ \|x\|_0 \leq 120$) and then set $y: = Ax^* . $    We perform the   HTP, ROTP, ROTP2 and ROTP3 up to 50 iterations, and the values of the residual $\|y-Ax^p\|_2$  with respect to the number of iterations for these algorithms are  described in Fig \ref{Fig-A} (a). It can clearly be  seen that our algorithms   are  stable in the sense that the  residual is successively reduced to the prescribed tolerance $\|y-Ax^p\|_2 \leq 10^{-8}   $ within a small number of iterations.
From Fig \ref{Fig-A} (a), however,  the  residuals at the iterates generated by the HTP oscillate dramatically with no clear movement towards the solution of the problem over  the course of iterations.   This oscillation phenomenon in hard thresholding pursuits was not observed in the ROTP and its enhanced versions, although such experiments were repeated a number of times on random examples of the problems.  This experiment also indicates that the number of iterations required by the ROTP2 and ROTP3 to find the solution of a  problem is  lower than the number of iterations required by the ROTP.  This means compressing the vector $u^p= x^p+ A^T (y-Ax^p) $ more than once does  improve the stability and efficiency of the algorithm, as predicted in Section 3.

The second experiment was performed to demonstrate the average number of iterations required by the proposed algorithms to meet a prescribed recovery criterion.       In this experiment, we set $n=1000$ and $m= \beta n, $ where the ratio $\beta= m/n$ is ranged from 0.1 to 0.6 with  stepsize 0.025.  For every such ratio,   a random $k$-sparse vector $x^*$ with  $ k= \lfloor m/10 \rfloor $  and 50 random matrices $A\in \mathbb{R}^{m\times n}  $ were generated.   We set  $y:= A x^* $ as the measurements of $x^* $ for every generated matrix $A. $ The maximum number of iterations was set to be 50 for all algorithms.   The average numbers of iterations required by the ROTP, ROTP2 and ROTP3 to meet the recovery criterion $ \| x^{p}- x^*\|/ \|x^*\|_2 \leq 10^{-2} $ are summarized in Fig. \ref{Fig-A} (b) which  shows that the ROTP3 need averagely a smaller  number of iterations than the ROPT2, and both need a smaller number of iterations than the ROTP to meet the recovery criterion. When the ratio is relatively high, all these algorithms only require  a small number of iterations to meet the criterion. However,  the average number of iterations required by these algorithms increases as the ratio $ m/n$ decreases. When the ratio $ m/n$ drops to a certain threshold,  the number of iterations required by the ROTP to meet the recovery criterion goes above and beyond the prescribed maximum number of iterations, and thus the algorithm terminates after 50 iterations.



\begin{figure} [htp]
 $ \begin{array}{c}
\includegraphics [width=0.45\textwidth,
totalheight=0.225\textheight] {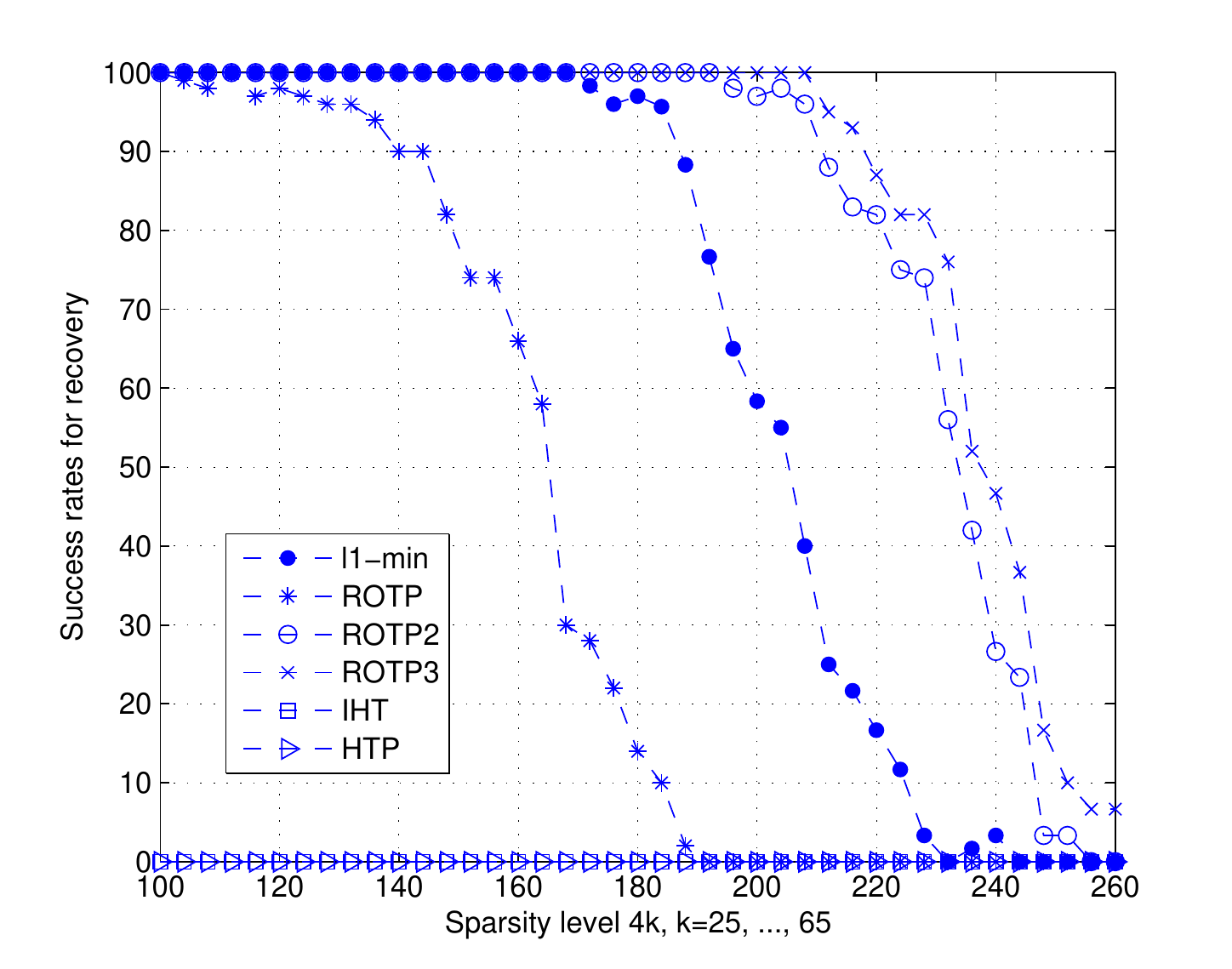} \\
  \textrm{(a) Sparse signals and inaccurate} \\
  \textrm{measurements}
\end{array} $
 \hfill   $\begin{array}{c}
\includegraphics [width=0.45\textwidth,
totalheight=0.225\textheight] {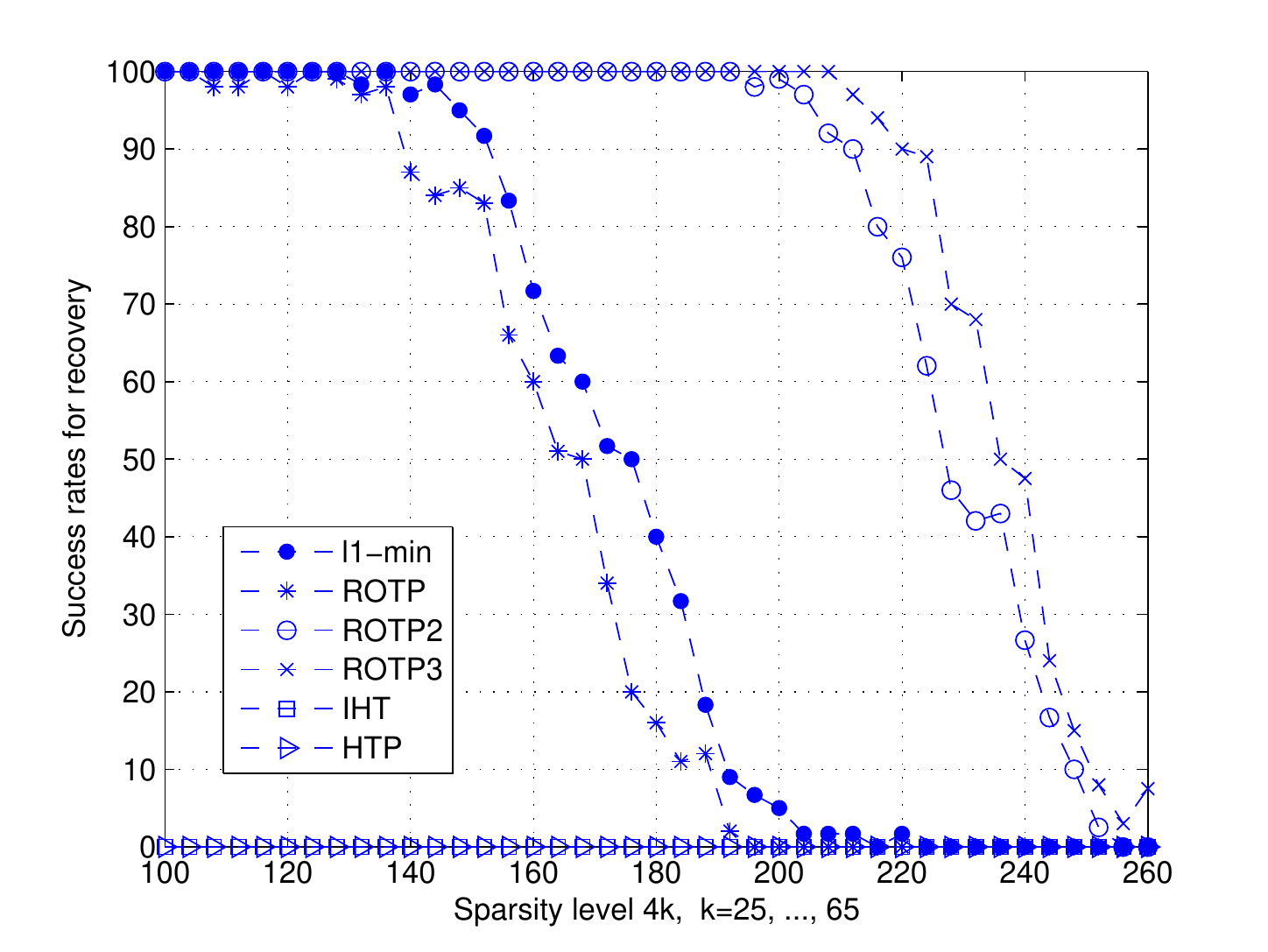} \\
\textrm{(b) Noisy signals and inaccurate  }\\
\textrm{measurements}
\end{array}
$
 \caption{Comparison of the success frequencies of the algorithms for signal recovery with inaccurate measurements. For every sparsity level, 50 random examples were realized. } \label{Fig-C}
\end{figure}

The other two experiments were carried out to compare our algorithms with several existing ones in terms of success frequencies of signal recovery.  The first comparison was done for the $k$-sparse  signal recovery with noisy measurements. The second comparison was done for both noisy signals and noisy measurements.
    We use the algorithms to recover, respectively, the sparse vectors $x^*\in \mathbb{R}^{1000}$ with different sparsity levels $\|x^*\|_0\leq 4 k,$ where $k=25, 26, \dots, 65, $ and their noisy counterparts  $\widetilde{x} $ which are approximately $k$-sparse.   For every such sparsity level,  we performed 50 random trials of the pair $(A, x^*), $ where $ A\in \mathbb{R}^{500 \times 1000}.  $ In the first comparison, we set $ y= Ax^* + \epsilon \theta$  as the measurements of $x^*,$ where $\epsilon =0.01$ and $\theta \in \mathbb{R}^n$ is a random  noise vector  with each component following a  ${\cal N} (0,1) $ distribution.
  We applied the IHT, HTP, $\ell_1$-minimization, ROTP, ROTP2 and ROTP3  to these  recovery problems, respectively, and we adopted $ \| x^{p}- x^*\|/ \|x^*\|_2 \leq 10^{-2}$ as the stopping criterion.
  When an iterate $x^p$ satisfies this criterion,  the algorithm terminates and  a ``success" is counted; otherwise an ``unsuccess" is counted.  If the above  criterion is not satisfied after the algorithm has been performed 50 iterations (which was set as the maximum number of iterations in our experiments), then the algorithm still terminates and an ``unsuccess" is counted.
In the second comparison, we generated $A$ by the same way as  the first comparison. The non-sparse vectors  $\widetilde{x} $ were generated by adding the noises to the sparse vectors $x^*,$ i.e.,   $\widetilde{x} =x^*+\widetilde{\epsilon} \widetilde{\theta} $ where $\widetilde{\epsilon} =0.001  $ and $\widetilde{\theta}$ is a random noise vector with each entry having a ${\cal N} (0,1)  $ distribution.   We then set $  y: = A\widetilde{x}  + \epsilon \theta $ as the measurements of $\widetilde{x},$ where  $\epsilon =0.01  $ and   $\theta$ is a random  noise vector with each entry following ${\cal N} (0,1). $  The stopping criterion for this case was chosen as $ \| x^{p}-  \widetilde{x}_S \|/ \| \widetilde{x}_S \|_2 \leq 10^{-2},$ where $ S $ is the index set for the $4k$ largest magnitudes of $\widetilde{x},$ where $k=25, \dots, 65.$
  The success rates of the  algorithm are summarized in Fig. \ref{Fig-C}, in which  (a) is the result for the case in which $y$ is inaccurate and $x^*$ is $k$-sparse, and  (b) is the result for both noisy   measurements and noisy signals. The experiments indicate that  the ROTP, ROTP2 and ROTP3  remarkably outperform  the traditional IHT and HTP that  fail to recover the vectors with sparsity in the above-mentioned ranges.   More interestingly, the ROTP2 and ROTP3 outperform the ROTP  and remarkably  outperform the $\ell_1$-minimization method,  especially in noise scenarios. The experiments   indicate that  the success rates of $\ell$-minimization is somewhat sensitive to the noise level of the signals. Our algorithms, however,  is more robust than $\ell$-minimization for noisy  signal recovery.

\section{Conclusions and future work}

The oscillation phenomenon in   hard thresholding pursuits can be overcome by linking the $k$-thresholding with   residual reductions. The optimal thresholding technique introduced in this paper naturally leads to the  relaxed optimal $k$-thresholding pursuit (ROTP) and its enhanced counterparts, ROTP2 and ROTP3, which turn out to be efficient numerical methods for sparse optimization problems.  The experiments indicate that the residual can be successively reduced in the course of iterations of the proposed algorithms, and thus the iterates generated by these algorithms move in a stable manner towards the solution of the sparse optimization problems.  The essential idea for this new development is that \emph{the hard thresholding operator should be applied to a compressible vector, instead of any vector.}  The OT and OTP provide a fundamental  basis for the development of such efficient numerical methods. Motivated by this study, several research directions are worthwhile to pursue in the near future. For instance, the recovery bound $\delta_{2k}\leq \tau^*$ in Theorem \ref {Thm-OIHT-01}  goes beyond the   bounds for traditional hard thresholding methods. However, this bound remains largely theoretical from the perspective that directly solving the binary quadratic optimization problem in OT or OTP remains challenging, especially in high-dimensional settings.  How to use the modern integer programming  techniques to deal with the subproblems in OT and OTP without relying on the convex relaxation technique is one of the interesting future  work. In addition, the study in this paper demonstrates that the ROPT,  ROPT2 and ROTP3 derived from convex relaxation are very efficient thresholding methods compared with existing ones.  However,  the first convergence result for the ROTP was shown in this paper under the condition $\delta_{3k} \leq 1/5 $ which is relatively restrictive.  Whether this result can be improved is also a worthwhile question to address in the near future. Moreover, the  optimal  thresholding technique  introduced in this paper can be  used to  stabilize any sparsity-seeking procedures provided that the hard thresholding operator is involved in the procedure, such as compressed sampling matching pursuits, subspace pursuits and the graded hard thresholding pursuits. So a  further development for these procedures can be  anticipated as well. We use this paper to develop a preliminary theory but a key step towards such a further development.


\begin{thebibliography}{999}



\bibitem{BE13} A. Beck and Y.C. Eldar,  Sparse signal recovery from nonlinear measurements,  ICASSP 2013, IEEE, pp. 5464--5468.

 \bibitem{BE13b}  A. Beck and Y.C. Eldar,  Sparsity constrained nonlinear optimization: Optimality conditions and algorithms, \emph{SIAM J. Optim.}, 23 (2013), pp. 1480--1509.

\bibitem{BT09} A. Beck and M. Teboulle,   A fast iterative shrinkage-thresholding algorithm for linear inverse problems, \emph{SIAM J. Imaging Sci.}, 2 (2009), pp. 183--202.

\bibitem{BKM16} D. Bertsimas, A. King and R. Mazumder, Best subset selection via a modern optimization Lens, \emph{Ann. Statist.}, 44 (2016), no.2, pp. 813--852.

\bibitem{BTW15} J.D. Blanchard, J. Tanner and  K. Wei, CGIHT: Conjugate gradient iterative hard thresholding for compressed sensing and matrix completion, \emph{IEEE Trans. Signal Process.}, 63 (2015), pp. 528-537.


 \bibitem{B12} T. Blumensath, Accelerated iterative hard thresholding, \emph{Signal Process.}, 92 (2012), 752--756.

\bibitem{BD08} T. Blumensath and M.E. Davies, Iterative hard thresholding for sparse approximation, \emph{J. Fourier Anal. Appl.},  14 (2008), pp. 629--654.

\bibitem{BD09} T. Blumensath and M.E. Davies, Iterative hard thresholding for compressed sensing, \emph{Appl. Comput. Harmon. Anal.}, 27 (2009), pp. 265--274.

\bibitem{BD10} T. Blumensath and M.E. Davies, Normalized iterative hard thresholding: Guaranteed stability and performance, \emph{IEEE J. Sel. Top. Signal Process.}, 4 (2010), pp. 298--309.

\bibitem{B14} J.-U. Bouchot,  A generalized class of hard thresholding algorithms for sparse signal recovery. In: Fasshauer G., Schumaker L. (eds) Approximation Theory XIV: San Antonio 2013.  Springer Proceedings in Mathematics \& Statistics, 83 (2014), pp. 45--63.


\bibitem{BFH16} J.-U., Bouchot, S. Foucart and P. Hitczenki,  Hard thresholding pursuit algorithms: Number of iterations,  \emph{Appl. Comput. Harmon. Anal.},  41 (2016), pp. 412-435.

\bibitem{BDE09} A.M. Bruckstein, D.L. Donoho and M. Elad,
From sparse solutions of systems of equations to sparse modeling of
signals and images, \emph{SIAM Rev.}, 51 (2009), pp. 34--81.

\bibitem{BT18} C. Buchheim
and E. Traversi, Quadratic combinatorial optimization using separable underestimators, \emph{INFORMS Journal on Computing}, 30 (2018),
pp. 424--637.


\bibitem{C08} E.J. Cand\`es, The restricted isometry property and its
implications for compressed sensing,  \emph{C.R. Math. Acad. Sci.
paris}, 346 (2008), pp. 589--592.

\bibitem{CT05} E.J. Cand$\grave{\textrm{e}}$s and T. Tao, Decoding by linear
programming, \emph{IEEE Trans. Inform. Theory}, 51 (2005), pp.
4203--4215.

\bibitem{CWB08} E.J. Cand$\grave{\textrm{e}}$s, M. Wakin and S.
Boyd, Enhancing sparsity by reweighted $\ell_1$ minimization,
\emph{J. Fourier Anal. Appl.}, 14 (2008), pp. 877--905.

\bibitem{C11} V.  Cevher, On accelerated hard thresholding methods for sparse approximation, Proc. SPIE 8138, Wavelets and Sparsity XIV, 813811, 2011.

\bibitem{CAP08} W.A. Chaovalitwongse, I.P. Androulakis and P.M. Pardalos,   Quadratic integer programming: Complexity and equivalent forms. In: Floudas C., Pardalos P. (eds) \emph{Encyclopedia of Optimization}, Springer, Boston, MA, 2008.

\bibitem{CDS98} S.S. Chen, D.L. Donoho and M.A. Saunders, Atomic
decomposition by basis pursuit, \emph{SIAM J. Sci. Comput.}, 20 (1998),
pp. 33--61.


\bibitem{DM09} W. Dai, and O. Milenkovic, Subspace pursuit for compressive sensing signal reconstruction, \emph{IEEE Trans. Inform. Theory}, 55 (2009), pp. 2230--2249.

\bibitem{DDM04} I. Daubechies, M. Defries and C. De Mol, An iterative thresholding algorithm for linear inverse problems with a sparsity constraint, \emph{Comm. Pure Appl. Math.}, 57 (2004), pp. 1413--1457.


\bibitem{D95} D.L. Donoho,  De-noising by soft-thresholdinng, \emph{IEEE Trans. Inform. Theory}, 41 (1995), pp. 613--627.

\bibitem{DJ94} D.L. Donoho and I. Johnstone, Idea spatial adaptation via wavelet shrinkage, \emph{Biomatrika}, 81 (1994), pp. 425--455.

\bibitem{E06} M. Elad, Why simple shringkage is still relevant for redundant representation,  \emph{IEEE Trans. Inform. Theory},  52 (2006),   pp. 5559--5569.

\bibitem{E10} M. Elad, \emph{Sparse and Redundant Representations: From Theory to Applications in Signal
and Image Processing}, Springer, New York, 2010.



\bibitem{EK12} Y.C. Eldar and G. Kutyniok, \emph{Compressed Sensing: Theory
and Applications}, Cambridge University Press, 2012.

\bibitem{FN03} M. Figueiredo and R. Nowak, An EM algorithm for wavelet-based image restoration, \emph{IEEE Trans. Image Process.}, 12 (2003), pp. 906--916.

\bibitem{FR08} M. Fornasier and R. Rauhut, Iterative thresholding algorithms, \emph{Appl. Comput. Harmon. Anal.}, 25 (2008), pp. 187-208.

 \bibitem{F12} S. Foucart,    Sparse recovery algorithms: Sufficient conditions in terms of restricted isometry constants. In: Neamtu M., Schumaker L. (eds) Approximation Theory XIII: San Antonio 2010. Springer Proceedings in Mathematics, 13 (2012),  pp. 65-77.

\bibitem{F11} S. Foucart, Hard thresholding pursuit: An algorithm for compressive sensing, \emph{SIAM J. Numer. Anal.}, 49 (2011), pp. 2543--2563.


\bibitem{FL09} S. Foucart an M. Lai,  Sparsed solutions of underdetermined linear systems via $\ell_q$-minimization for $0\leq q\leq 1,$ \emph{Appl. Comput. Harmon. Anal.}, 26 (2009), pp. 395--407.

\bibitem{FR13} S. Foucart and H. Rauhut, \emph{A Mathematical Introduction to Compressive Sensing},
  Springer, NY, 2013.

  \bibitem{GK09} R. Garg and	R.  Khandekar,
Gradient descent with sparsification: An iterative algorithm for sparse recovery with restricted isometry property, Proceeding
ICML 2009,  Montreal,  Canada, pp. 337-344.

\bibitem{GB11} M. Grant and S. Boyd,
\emph{CVX: Matlab Software for Disciplined Convex Programming},
Version 1.21, April 2017.

\bibitem{HGT06} K. Herrity, A.  Gilbert and J. Tropp,  Sparse approximation via iterative thresholding, in IEEE ICASSP 2006, pp. 624--627.

\bibitem{L51} L. Landweber, An iteration formula for Freholm integral equations of the first kind, \emph{Amer. J. Math.}, 73 (1951), pp. 615--624.

\bibitem{L16} K. Lange,  \emph{MM Optimization Algorithms},  SIAM, Philadelphia, 2016.


\bibitem{KK17} R. Khanna, and A. Kyrillidis, IHT dies hard:  Provable accelerated iterative hard thresholding, Preprint,  2017.

\bibitem{KR03} N. Kingsbury and T. Reeves,  Redundant representation with complex wavelets:  How to achieve sparsity,  in IEEE ICIP 2003, Barcelona,  pp. 45--48.

\bibitem{KC14}  A. Kyrillidis and V. Cevher, Matrix recipes for hard thresholding methods,  \emph{J. Math. Imag. Vision},  48 (2014), pp. 235--265.

\bibitem{LYSM17} H. Liu, M.-C. Yue, A.M.-C. So and W.-K., Ma,  A discrete first-order method for large-scale MIMO detection with provable guarantees, in \emph{Proc. of IEEE 18th  Internal Workshop on SPAWC}, 2017.

\bibitem{M09} A. Maleki, Coherence analysis of iterative thresholding algorithms, Forty-Seventh Annual Allerton Conference
Allerton House, UIUC, Illinois, USA, 2009, pp. 236--243.


\bibitem{MZ93} S. Mallat and Z. Zhang, Matching pursuits with
time-frequency dictionaries, \emph{IEEE Trans. Signal Process.}, 41
(1993), pp. 3397--3415.

\bibitem{MDZ94} S. Mallat, G. Davis and Z. Zhang, Adaptive time-frequency decompositions, \emph{SPIE J. Opt. Eng.}, 33 (7), (1994), pp. 2183--2191.

\bibitem{M02} A. Miller, \emph{Subset Selection in Regression},  CRC Press,  Washington, 2002.

\bibitem{N95} B.K. Natarajan, Sparse approximate solutions to linear
systems, \emph{SIAM J. Comput.}, 24 (1995), pp. 227-234.

\bibitem{NT09} D. Needell and J.A. Tropp,  CoSaMP:  Iterative signal recovery from incomplete and inaccurate samples, \emph{Appl. Comput. Harmon. Anal.} 26 (2009), pp. 301--321.

\bibitem{N13} Y. Nesterov, \emph{Introductory Lectures on Convex Optimization: A Basic Course}, Volume 87, Springer Science and Business Media, 2013.

\bibitem{RK02} T.H. Reeves and N.G. Kingsbury,  Overcomplete image coding using iterative projection-based noise shaping, in IEEE ICIP 2002, Rochester,    pp. 597--600.

\bibitem{SNM03} J. Starck, M. Nguyen, and  F. Murtagh, Wavelet and curvelet for image deconvolution: A combined approach, \emph{J. Signal Process.}, 83 (2003), pp. 2279--2283.

\bibitem{VW13}  S. Voronin, H.J. Woerdeman,  A new iterative firm-thresholding algorithms for inverse problems with sparsity constraints,  \emph{Appl. Comput. Harmonic Anal.}, 35 (2013), pp. 151--164.






\bibitem{TG07} J.A. Tropp and A.C. Gilbert, Signal recovery from
random measurements via orthogonal mathcing pursuit, \emph{IEEE
Trans. Inform. Theory}, 53 (2007), pp. 4655--4666.

\bibitem{Z18} Y.-B. Zhao, \emph{Sparse Optimization Theory and Methods}, CRC Press, Taylor \& Francis Group, Boca Raton, FL, 2018.

\bibitem{ZK15} Y.-B. Zhao and M. Ko{\v c}vara, A new computational  method for the sparsest solutions to
  systems of linear equations, \emph{SIAM J. Optim.},    25 (2015),  pp. 1110--1134.

\bibitem{ZL17} Y.-B. Zhao and Z.-Q. Luo,
Constructing new reweighted $\ell_1$-algorithms for the sparsest
 points of polyhedral sets, \emph{Math.  Oper.  Res.}, 42 (2017), pp. 57--76.

\bibitem{ZL12} Y.-B. Zhao and D. Li,   Reweighted $\ell_1$-minimization for sparse solutions to
 underdetermined linear systems, \emph{SIAM J. Optim.}, 22 (2012), pp. 893--912.

\end{thebibliography}
\end{document}